\documentclass[a4paper,conference]{IEEEtran}
\IEEEoverridecommandlockouts
\usepackage{amsmath}
\usepackage{amssymb}
\usepackage{mathrsfs}
\usepackage{bm}
\usepackage{verbatim}
\usepackage{array}
\usepackage{multirow}
\usepackage{subfigure}
\usepackage{caption}
\usepackage{algorithm}
\usepackage{algorithmicx}
\usepackage{algpseudocode}
\usepackage{makecell}
\usepackage{tabularx} 
\usepackage{booktabs}
\usepackage{rotating}
\usepackage{graphicx}
\usepackage{textcomp}
\begin{document}
\title{Snapshot Hyperspectral Imaging Based on Weighted High-order Singular Value Regularization}

%\author{
%	\IEEEauthorblockN{Hua Huang \quad Niankai Cheng\thanks{aas} \quad Lizhi Wang} 	
%	\IEEEauthorblockA{School of Computer Science and Technology, Beijing Institute of Technology} 
%	\IEEEauthorblockA{\{huahuang,naikai,lzwang\}@bit.edu.cn} 
%}

\author{\IEEEauthorblockN{Niankai Cheng\IEEEauthorrefmark{1},
		Hua Huang\thanks{Corresponding author: Hua Huang(huahuang@bnu.edu.cn)}\IEEEauthorrefmark{2}, Lei Zhang\IEEEauthorrefmark{1},
		and Lizhi Wang\IEEEauthorrefmark{1}}
	\IEEEauthorblockA{\IEEEauthorrefmark{1}School of Computer Science and Technology, Beijing Institute of Technology, Beijing 100081, China}
	\IEEEauthorblockA{\IEEEauthorrefmark{2}School of Artifical Intelligence, Beijing Normal University, Beijing 100875, China}}

\maketitle

\def\FigFrame{
	\begin{figure}[t]
		\begin{center}
			\includegraphics[width=0.90\columnwidth]{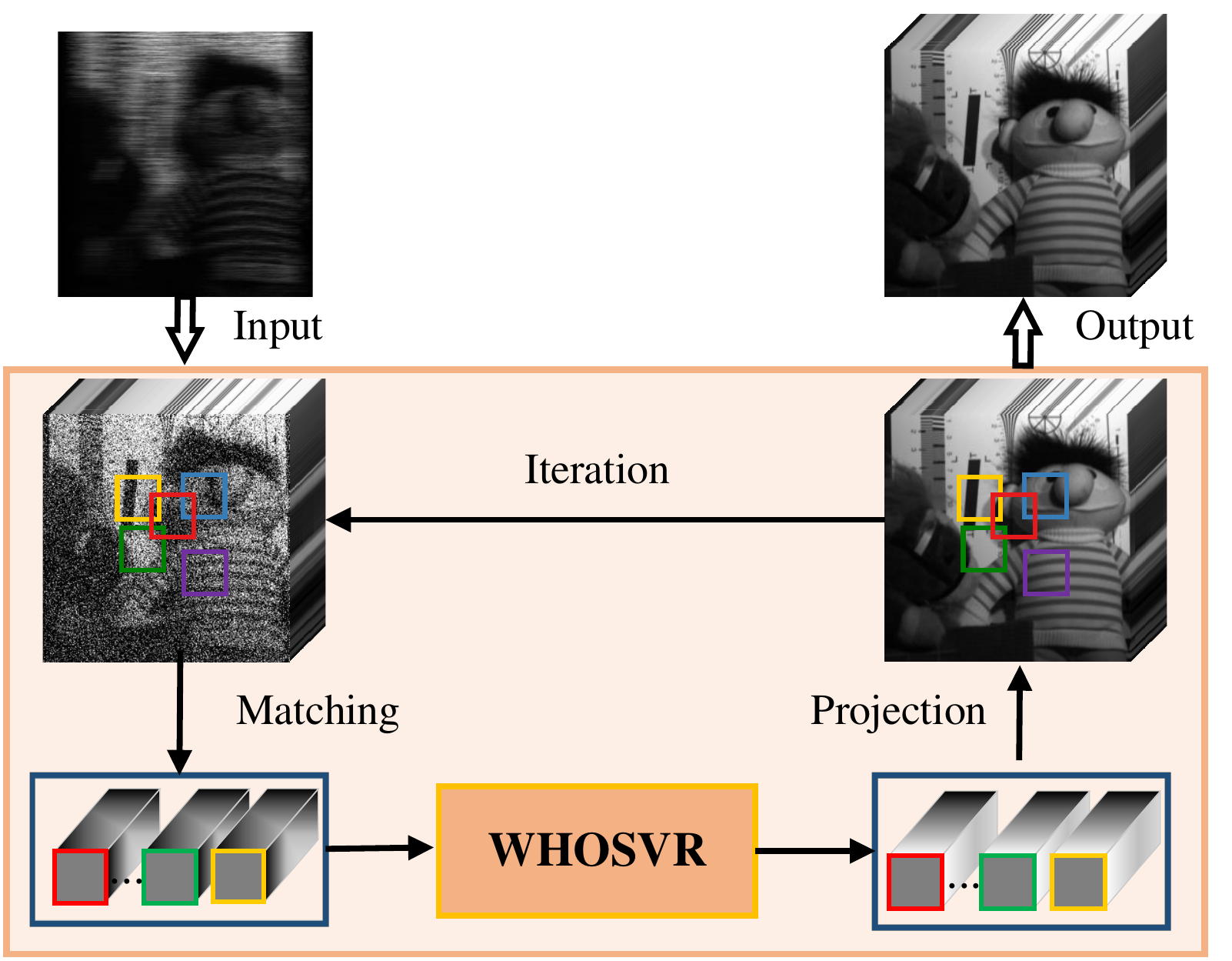}
		\end{center}
		%\vspace{-0.3cm}
		\caption{Overview of the proposed method. We reconstruct HSI from the compressive measurement. Our reconstruction method, including matching, weighted high-order singular value regularization (shortened as WHOSVR in the figure) and projection, is iteratively performed. }
		%\vspace{-0.4cm}
		\label{fig:System}
	\end{figure}
}

\def\FigTSVD{
	\begin{figure}[t]
		\begin{center}
			\includegraphics[width=1.0\columnwidth]{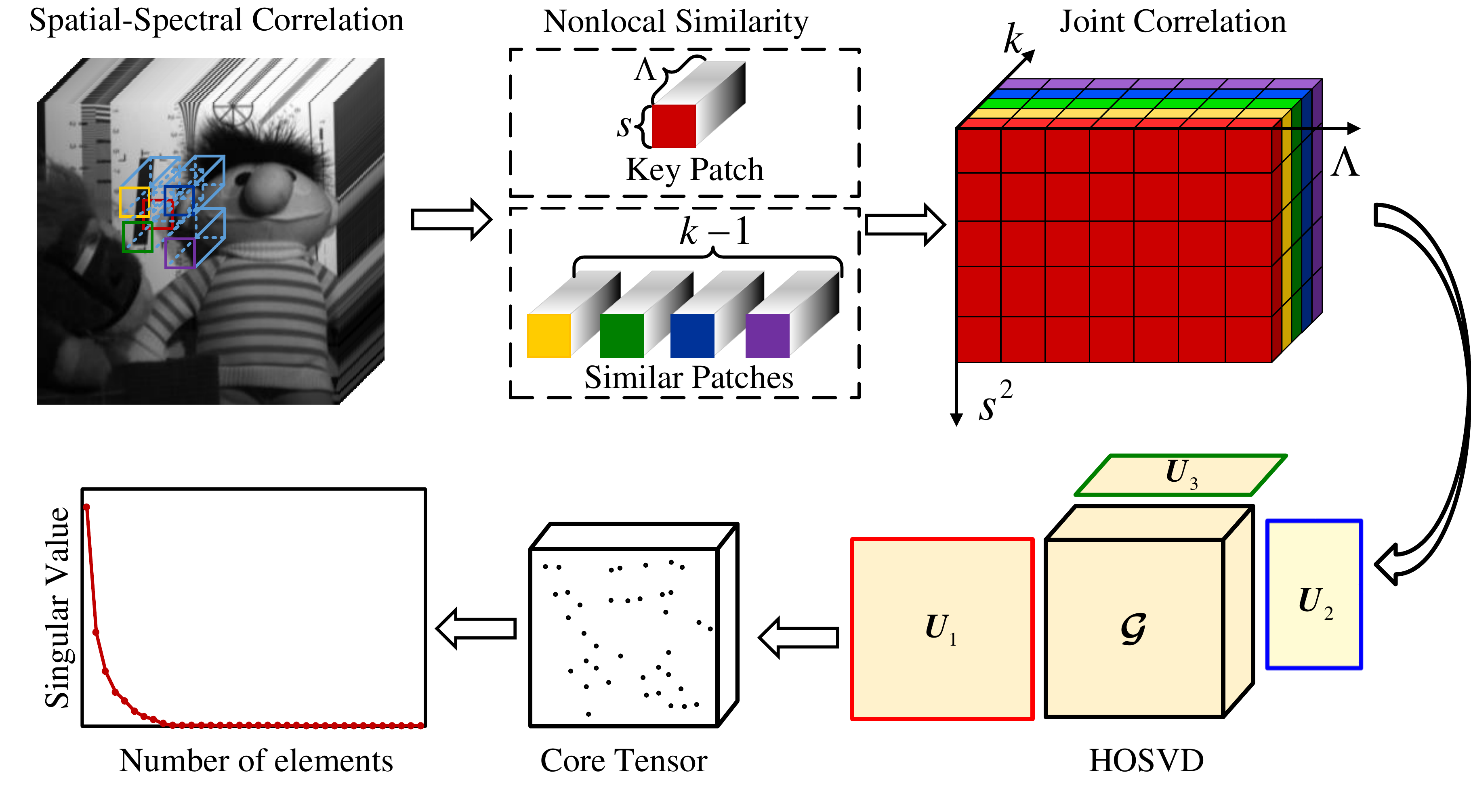}
		\end{center}
		%\vspace{-0.3cm}
		\caption{Low-rank property analysis. We exploit the nonlocal similarity across spatial and spectral dimensions to reformulate a low-rank tensor. Then we implement HOSVD on the tensor and show the distribution of singular values in the core tensor.}
		\label{fig:SVD}
		%\vspace{-0.4cm}
	\end{figure}
}

\def\FigSystem{
	\begin{figure}[t]
		\begin{center}
			\includegraphics[width=1.0\columnwidth]{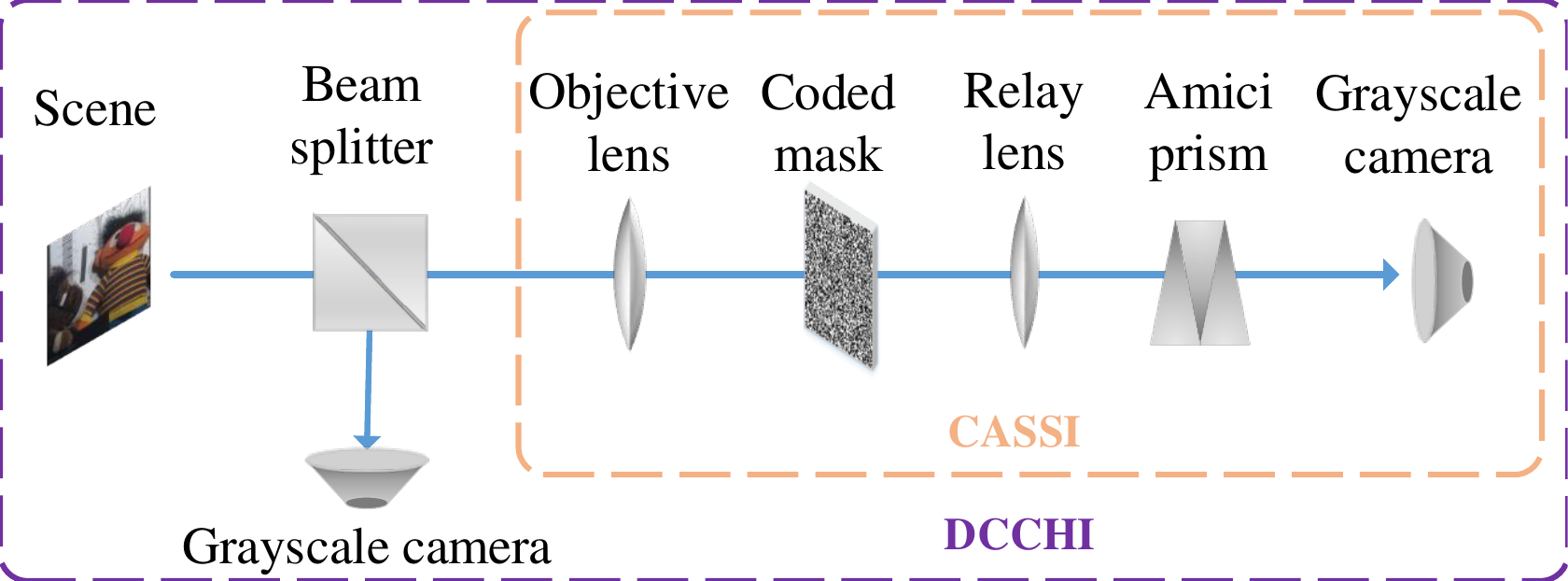}
		\end{center}
		%\vspace{-0.3cm}
		\caption{Diagram of two representative snapshot hyperspectral imaging systems.}
		%	\vspace{-0.4cm}
		\label{fig:DualCamera}
	\end{figure}
}

\def\TabSynCASSI{
	\renewcommand{\arraystretch}{1.5}
	% Table generated by Excel2LaTeX from sheet 'Sheet1'
	\begin{table*}[htbp]
		\centering
		\caption{Average reconstruction results (PSNR(dB)/SSIM/ERGAS/RMSE) of different methods on CASSI.}
		%	\vspace{-0.3cm}
		\begin{small}
			\begin{tabular}{|c|c|c|c|c|c|c|c|c|c|c|c|}
				\hline
				Indexes & TV    & AMP   & 3DSR  & NSR   & LRMA  & AE    & ISTA  & HSCNN & HRNet & SPR   & Ours \\
				\hline
				PSNR  & 23.16 & 23.18 & 23.636 & 26.13 & 25.94 & 25.72 & 20.60 & 25.09 & 22.83 & 24.48 & \textbf{28.05} \\
				\hline
				SSIM  & 0.7130 & 0.6600 & 0.7311 & 0.7610 & 0.7930 & 0.7720 & 0.5499 & 0.7334 & 0.6648 & 0.7395 & \textbf{0.8302} \\
				\hline
				ERGAS & 258.32 & 256.76 & 245.153 & 189.19 & 195.63 & 197.32 & 344.57 & 206.97 & 268.65 & 224.19 & \textbf{153.06} \\
				\hline
				RMSE  & 0.0469 & 0.0474 & 0.0457 & 0.0333 & 0.0315 & 0.0333 & 0.0653 & 0.0373 & 0.0496 & 0.0451 & \textbf{0.0236} \\
				\hline
			\end{tabular}%
		\end{small}
		
		\label{tab:CASSI}%
	\end{table*}%
}

\def\TabSynDCCHI{
	\renewcommand{\arraystretch}{1.6}
	
	\begin{table*}[t]
		\centering
		\caption{Average reconstruction results of different methods on DCCHI.}
		%	\vspace{-0.3cm}	
		\begin{small}
			\begin{tabular}{|c|c|c|c|c|c|c|}
				\hline
				Indexes & TV    & AMP   & 3DSR  & NSR  & LRMA  & Ours \\
				\hline
				PSNR  & 28.51 & 28.52 & 28.32 & 32.58 & 37.45 & \textbf{37.81} \\
				\hline
				SSIM  & 0.8938 & 0.8526 & 0.9037 & 0.9377 & 0.9730 & \textbf{0.9733} \\
				\hline
				ERGAS & 167.14 & 140.75 & 163.67 & 107.71 & 57.30 & \textbf{51.38} \\
				\hline
				RMSE  & 0.0525 & 0.0263 & 0.0337 & 0.0285 & 0.0132 & \textbf{0.0069} \\
				\hline
			\end{tabular}%
		\end{small}
		\label{tab:DCCHI}%
		%	\vspace{-0.5cm}	
	\end{table*}%
	%\end{small}
}

\def\TabSynCAVECASSI{
	\begin{table*}[htbp]
		\centering
		\renewcommand{\arraystretch}{1.4}
		\setlength{\abovecaptionskip}{0cm}
		\setlength{\belowcaptionskip}{0cm}
		\caption{Reconstruction results (PSNR(dB)/SSIM/ERGAS) of the 10 HSIs for different methods on DCD.}
		\setlength{\tabcolsep}{1.42mm}{
			\begin{tabular}{|c|c|c|c|c|c|c|c|c|c|c|c|c|}
				\hline
				Methods & Indexes & \includegraphics[width=0.35in]{graphics/rgb/toy_rgb.pdf}     & \includegraphics[width=0.35in]{graphics/rgb/cloth_rgb.pdf}     & \includegraphics[width=0.35in]{graphics/rgb/egyptian_rgb.pdf}     & \includegraphics[width=0.35in]{graphics/rgb/feathers_rgb.pdf}     & \includegraphics[width=0.35in]{graphics/rgb/flowers_rgb.pdf}     & \includegraphics[width=0.35in]{graphics/rgb/glass_rgb.pdf}     & \includegraphics[width=0.35in]{graphics/rgb/oil_rgb.pdf}     & \includegraphics[width=0.35in]{graphics/rgb/paints_rgb.pdf}     & \includegraphics[width=0.35in]{graphics/rgb/stuffed_rgb.pdf}     & \includegraphics[width=0.35in]{graphics/rgb/superballs_rgb.pdf}    & Average \\
				\hline
				\multirow{1.5}[6]{*}{TV\cite{Wagadarikar2008Single}} & PSNR  & 29.25 & 21.17 & 47.19 & 26.84 & 22.73 & 25.54 & 27.15 & 30.08 & 26.87 & 28.28 & 28.51 \\
				\cline{2-13}          & SSIM  & 0.921 & 0.793 & 0.994 & 0.912 & 0.822 & 0.880 & 0.873 & 0.854 & 0.956 & 0.934 & 0.894 \\
				\cline{2-13}          & ERGAS & 156.67 & 250.06 & 38.57 & 155.41 & 217.18 & 188.39 & 150.89 & 126.06 & 142.41 & 245.79 & 167.14 \\
				\hline
				\multirow{1.5}[6]{*}{GPSR\cite{Figueiredo2007Gradient}} & PSNR  & 27.62 & 25.99 & 36.97 & 25.86 & 24.69 & 25.13 & 27.88 & 29.48 & 26.47 & 27.92 & 27.80 \\
				\cline{2-13}         & SSIM  & 0.846 & 0.866 & 0.937 & 0.855 & 0.833 & 0.851 & 0.779 & 0.851 & 0.929 & 0.892 & 0.864 \\
				\cline{2-13}          & ERGAS & 181.06 & 133.41 & 140.06 & 164.91 & 172.74 & 194.91 & 138.77 & 144.54 & 151.69 & 241.40 & 166.35 \\
				\hline
				\multirow{1.5}[6]{*}{AMP\cite{7328255}} & PSNR  & 27.66 & 23.52 & 38.77 & 28.92 & 26.53 & 24.97 & 27.38 & 26.44 & 28.38 & 32.62 & 28.52 \\
				\cline{2-13}          & SSIM  & 0.829 & 0.793 & 0.955 & 0.880 & 0.841 & 0.819 & 0.882 & 0.760 & 0.869 & 0.897 & 0.853 \\
				\cline{2-13}          & ERGAS & 137.55 & 174.13 & 102.23 & 119.43 & 137.27 & 197.58 & 139.14 & 163.35 & 111.78 & 125.03 & 140.75 \\
				\hline
				\multirow{1.5}[6]{*}{NSR\cite{7676344}} & PSNR  & 31.01 & 22.27 & 48.49 & 36.77 & 30.87 & 27.21 & 28.02 & 35.45 & 30.11 & 35.58 & 32.58 \\
				\cline{2-13}          & SSIM  & 0.944 & 0.831 & \textbf{0.996} & 0.976 & 0.943 & 0.904 & 0.966 & 0.884 & 0.973 & 0.959 & 0.938 \\
				\cline{2-13}          & ERGAS & 135.46 & 223.83 & 33.61 & 47.36 & 85.87 & 156.51 & 137.69 & 59.04 & 98.78 & 98.90 & 107.71 \\
				\hline
				\multirow{1.5}[6]{*}{LRMA\cite{fu2016exploiting}} & PSNR  & 40.65 & 28.87 & 48.20 & 40.11 & 34.31 & 33.85 & 34.71 & 41.22 & 37.57 & 35.03 & 37.45 \\
				\cline{2-13}          & SSIM  & 0.989 & 0.927 & 0.991 & 0.989 & 0.971 & 0.970 & 0.973 & 0.945 & 0.991 & 0.986 & 0.973 \\
				\cline{2-13}          & ERGAS & 34.49 & 99.71 & 35.34 & 32.24 & 57.01 & 72.43 & 59.72 & 30.35 & 40.41 & 111.30 & 57.30 \\
				\hline
				\multirow{1.5}[6]{*}{Ours} & PSNR  & \textbf{42.02} & \textbf{29.79} & \textbf{48.91} & \textbf{40.67} & \textbf{35.37} & \textbf{34.80} & \textbf{35.27} & \textbf{41.94} & \textbf{38.30} & \textbf{35.88} & \textbf{38.29} \\
				\cline{2-13}          & SSIM  & \textbf{0.991} & \textbf{0.938} & 0.991 & \textbf{0.990} & \textbf{0.976} & \textbf{0.976} & \textbf{0.977} & \textbf{0.950} & \textbf{0.992} & \textbf{0.988} & \textbf{0.977} \\
				\cline{2-13}          & ERGAS & \textbf{25.98} & \textbf{89.58} & \textbf{33.08} & \textbf{30.13} & \textbf{50.36} & \textbf{64.68} & \textbf{55.79} & \textbf{27.90} & \textbf{37.06} & \textbf{103.58} & \textbf{51.81} \\
				\hline
			\end{tabular}%
		}
		\label{tab:DCDResults}%
	\end{table*}%
}

\def\SynCASSI{
	\renewcommand\arraystretch{1.1}
	\begin{figure*}[htbp] 
		\newcommand{\widthfigure}{0.15\linewidth}
		\newcommand{\nullspace}{0.001pt}
		\centering
		\setlength\tabcolsep{1pt}
		\begin{tabular}{cccccc}
			\includegraphics[width=\widthfigure]{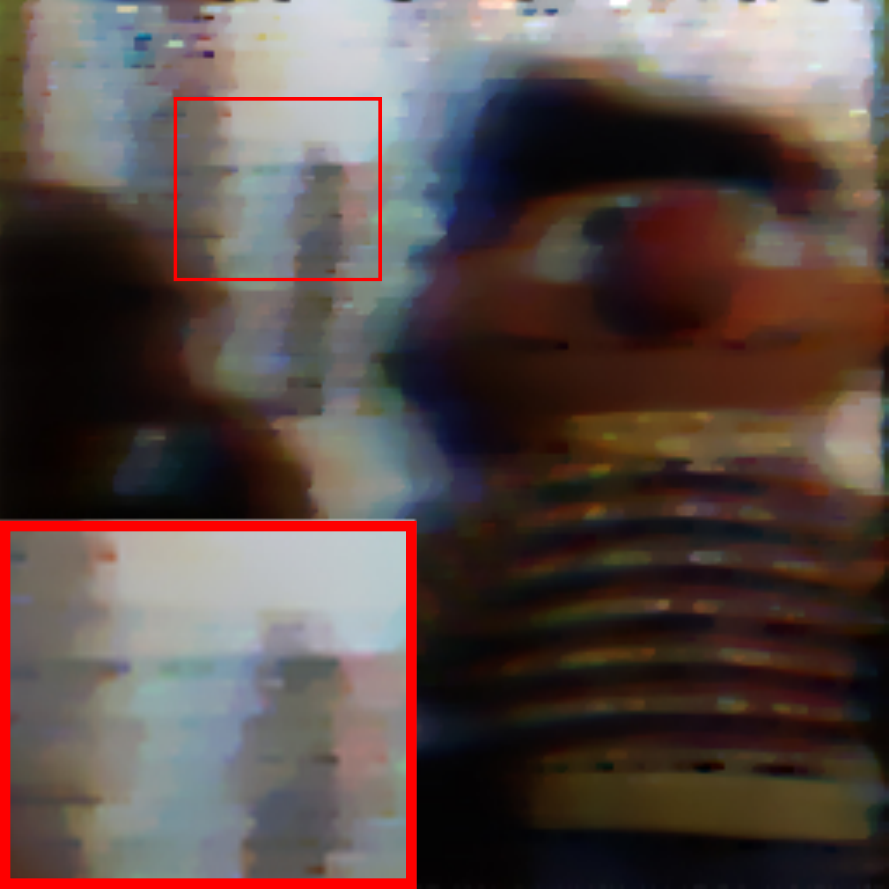}
			&
			\includegraphics[width=\widthfigure]{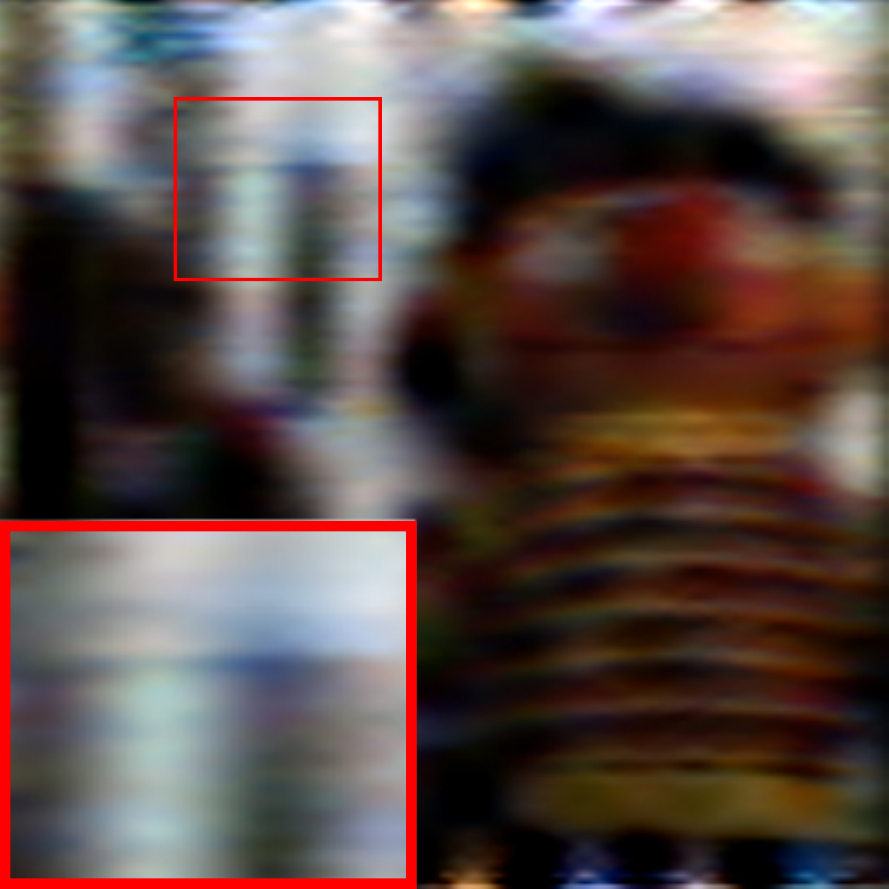}
			&
			\includegraphics[width=\widthfigure]{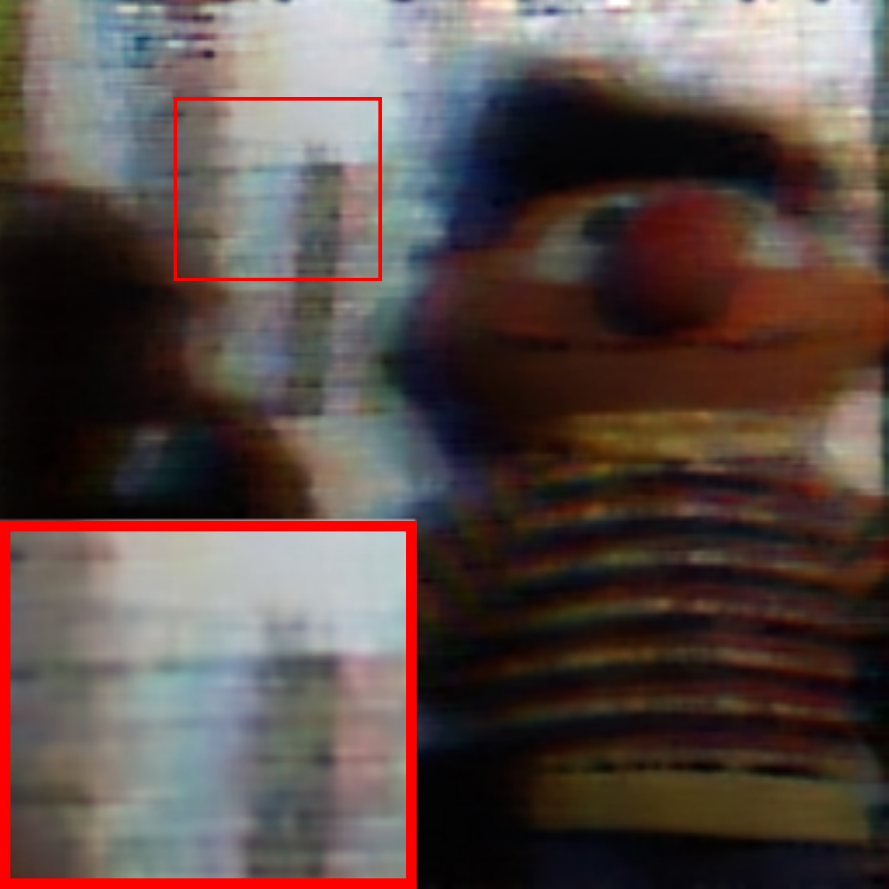}
			&
			\includegraphics[width=\widthfigure]{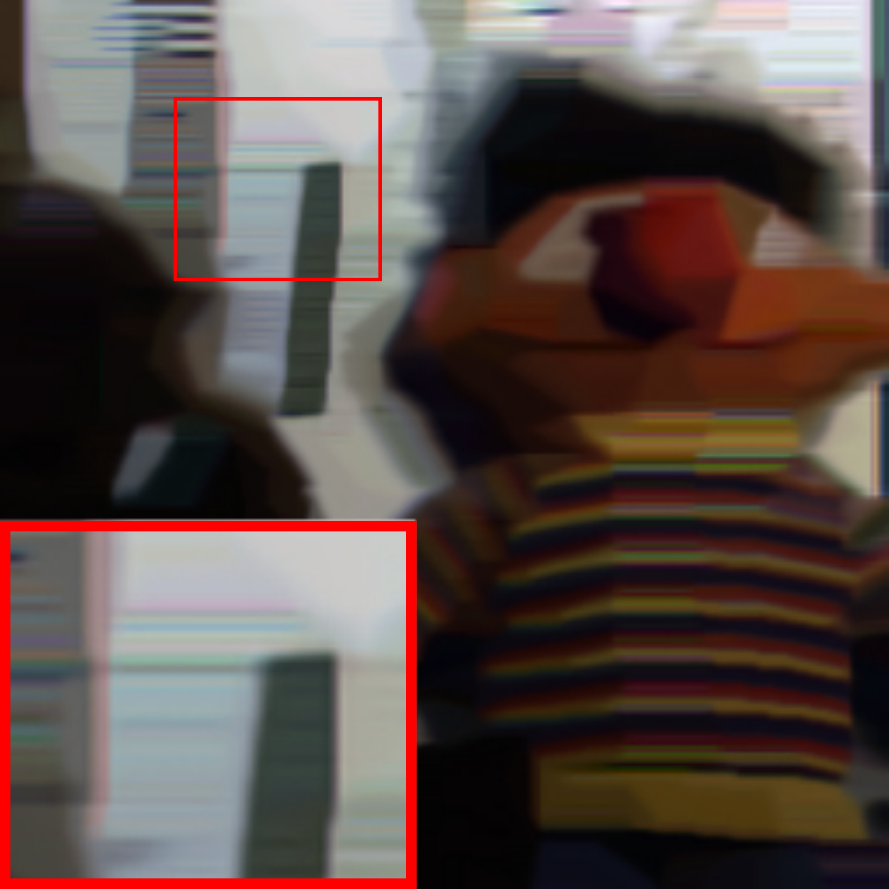}
			&
			\includegraphics[width=\widthfigure]{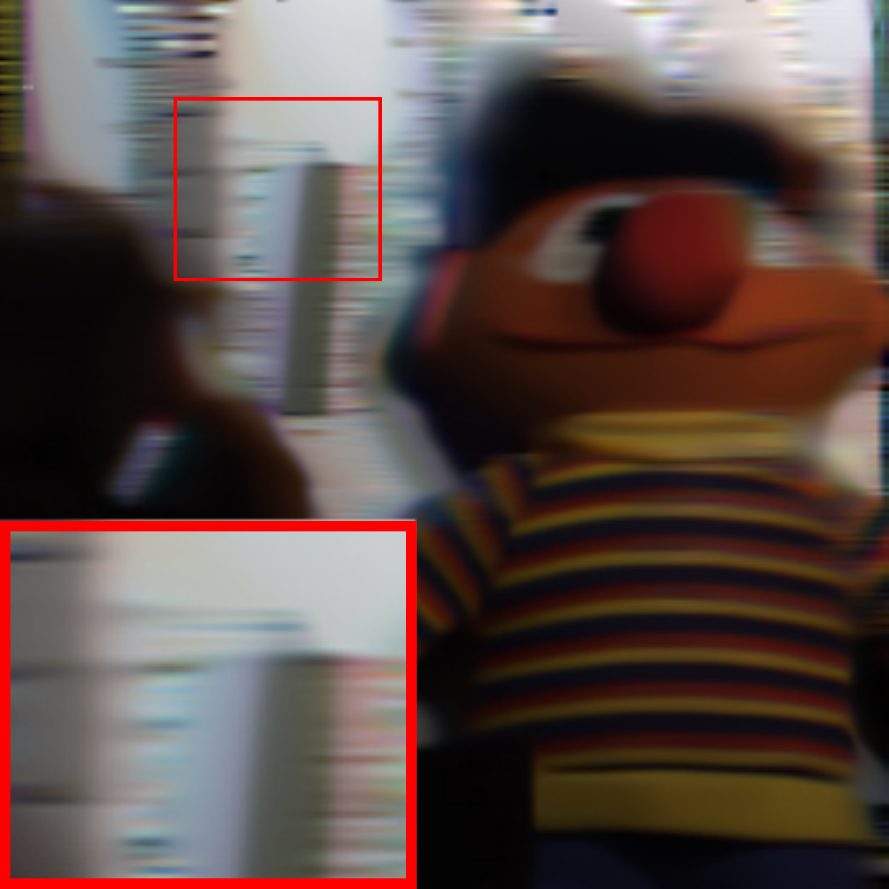}
			&
			\includegraphics[width=\widthfigure]{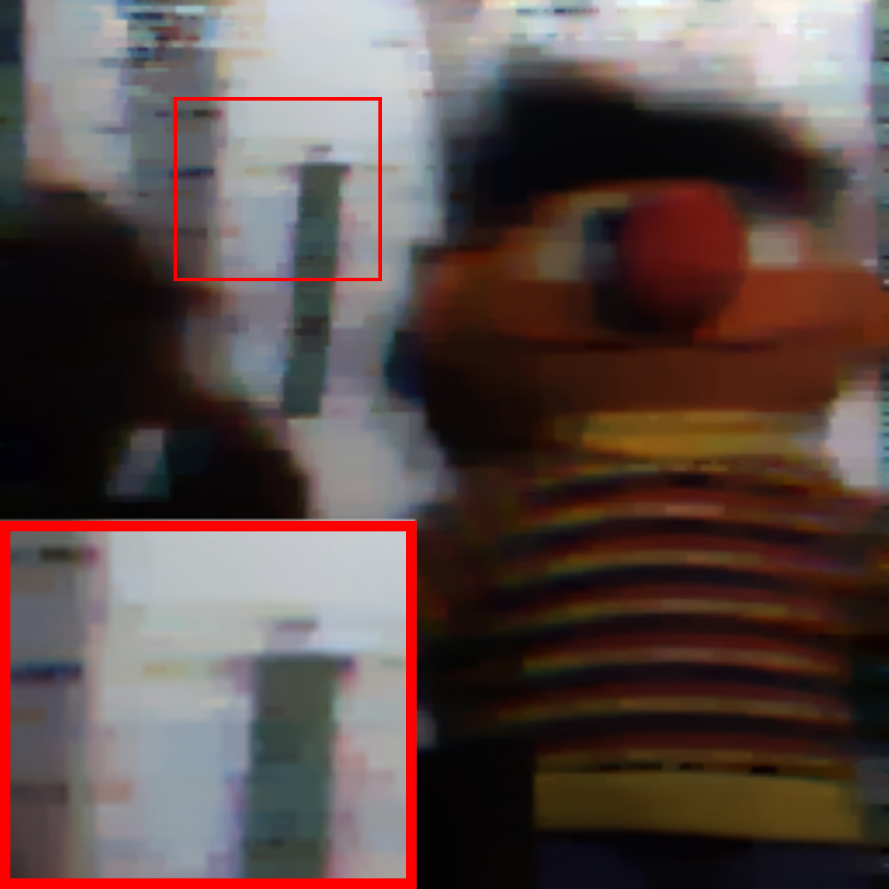}
			\\ TV & AMP & 3DSR & NSR & LRMA& AE\\
			(20.13 / 0.6750) & (19.61 / 0.5699) &(20.81 / 0.7030) & (23.46 / 0.7559) & (22.73 / 0.8063)& (22.84 / 0.7593)\\
			\\
			
			%		\midrule[0.7pt]
			\includegraphics[width=\widthfigure]{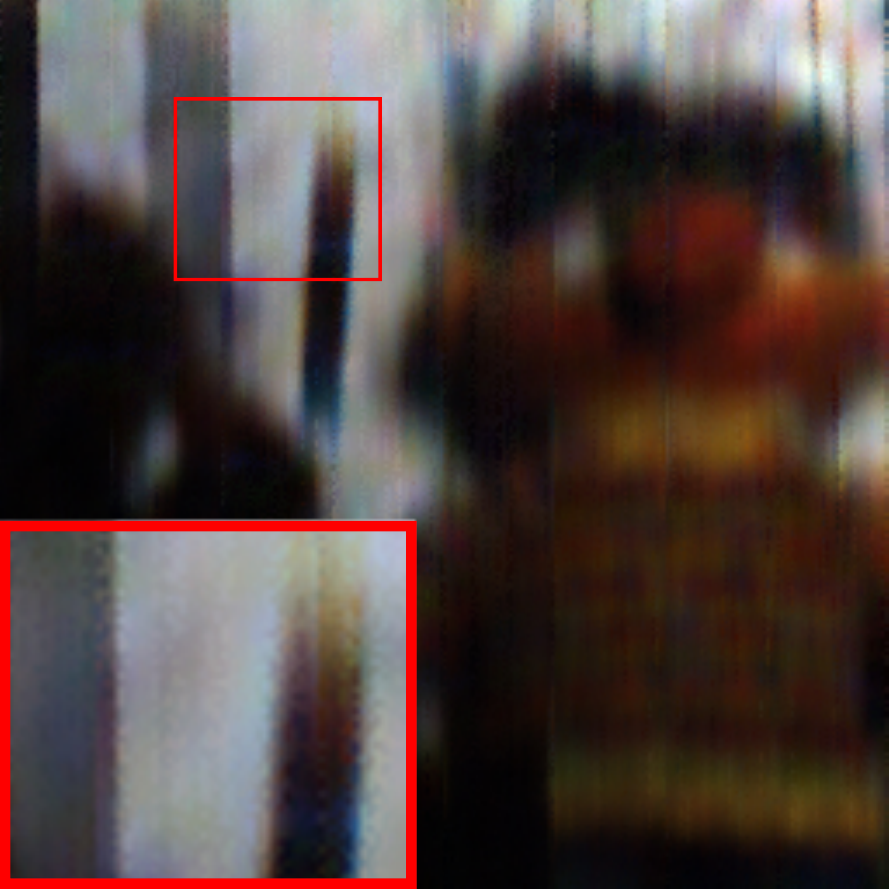}
			&
			\includegraphics[width=\widthfigure]{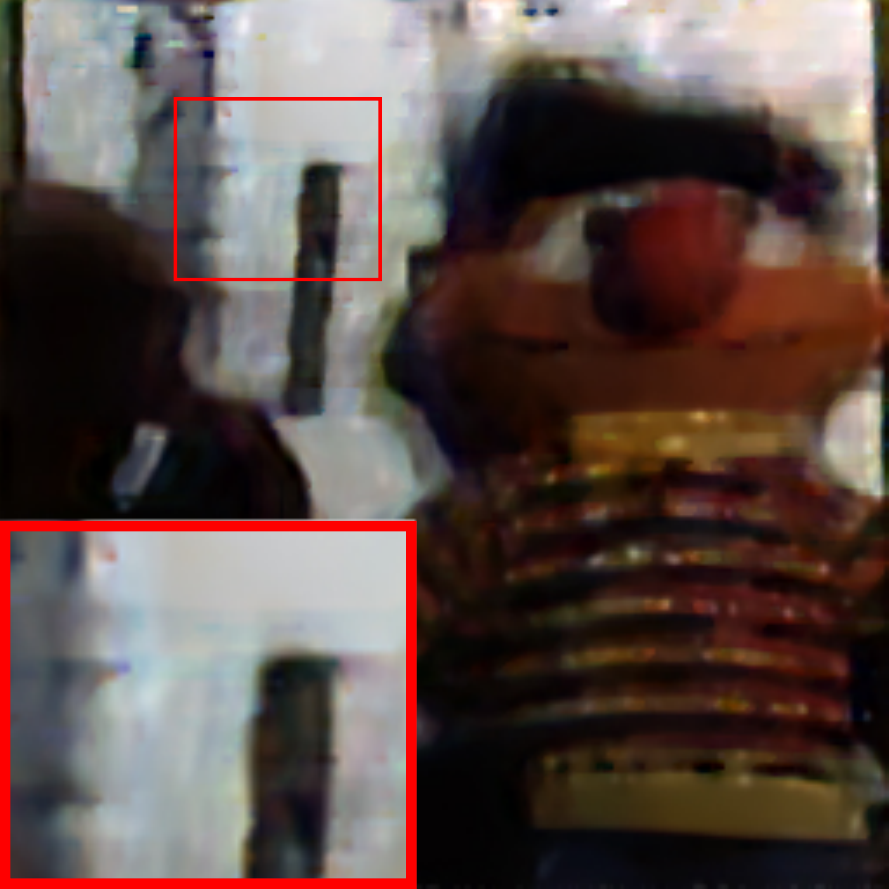}
			&
			\includegraphics[width=\widthfigure]{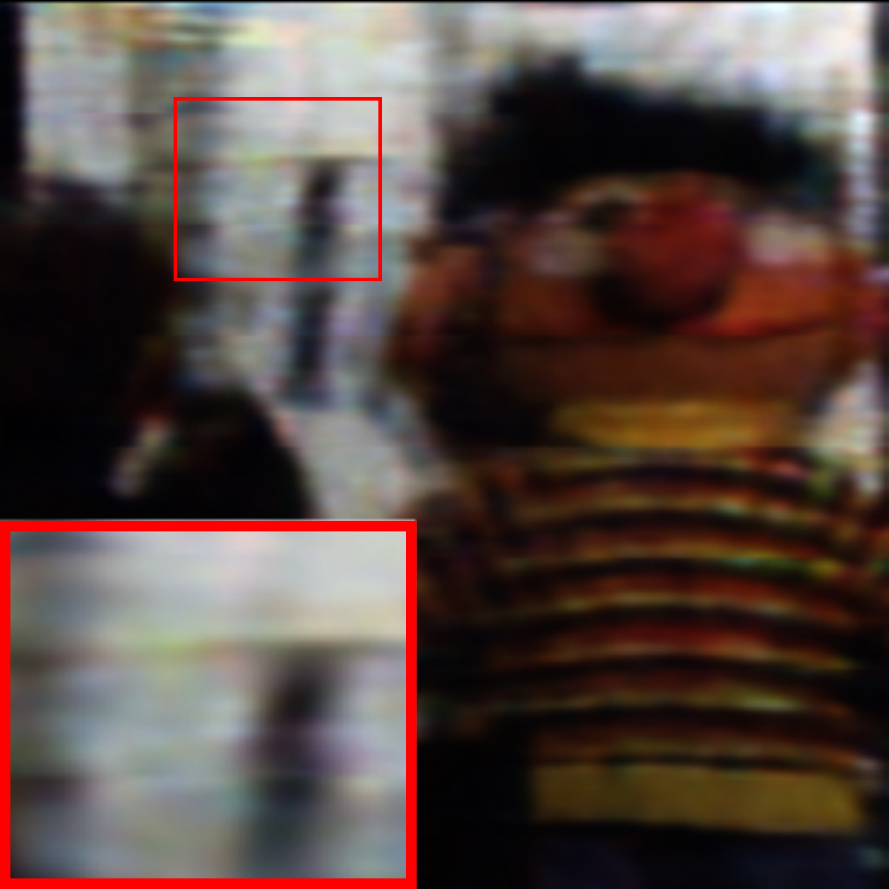}
			&
			\includegraphics[width=\widthfigure]{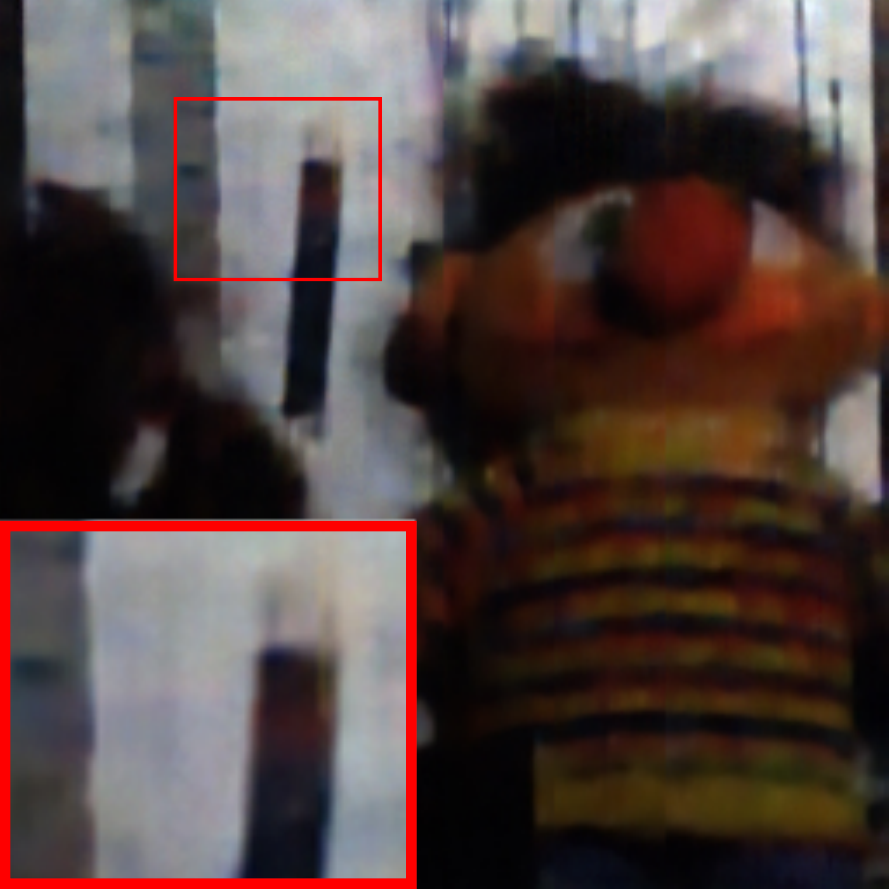}
			&
			\includegraphics[width=\widthfigure]{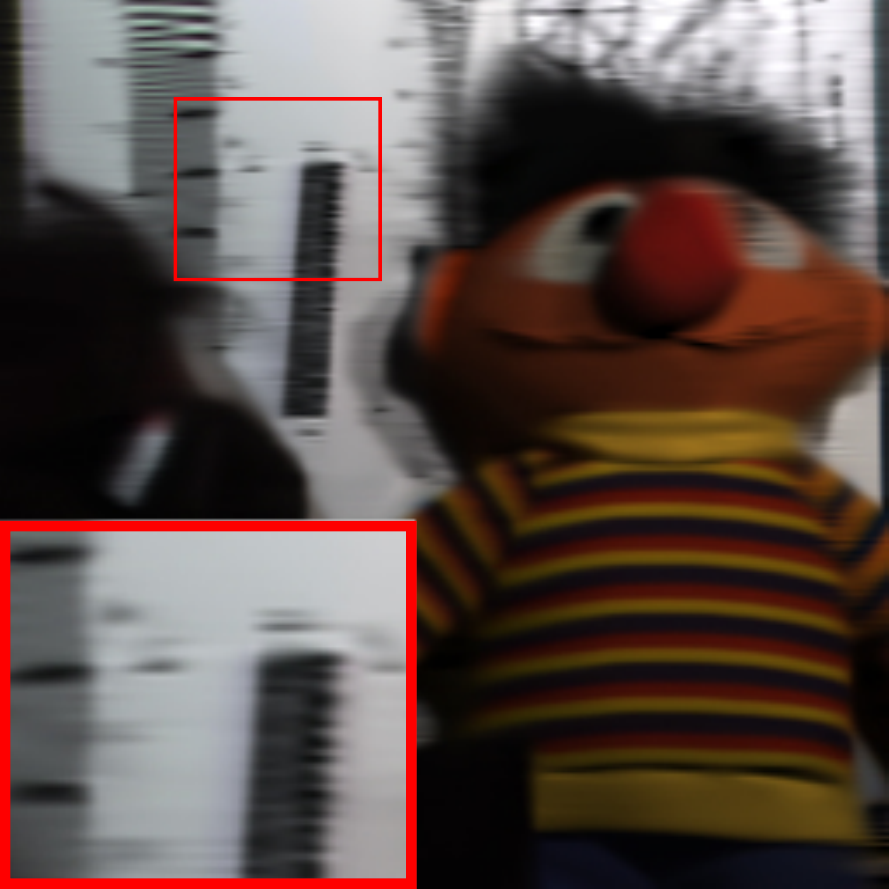}
			&
			\includegraphics[width=\widthfigure]{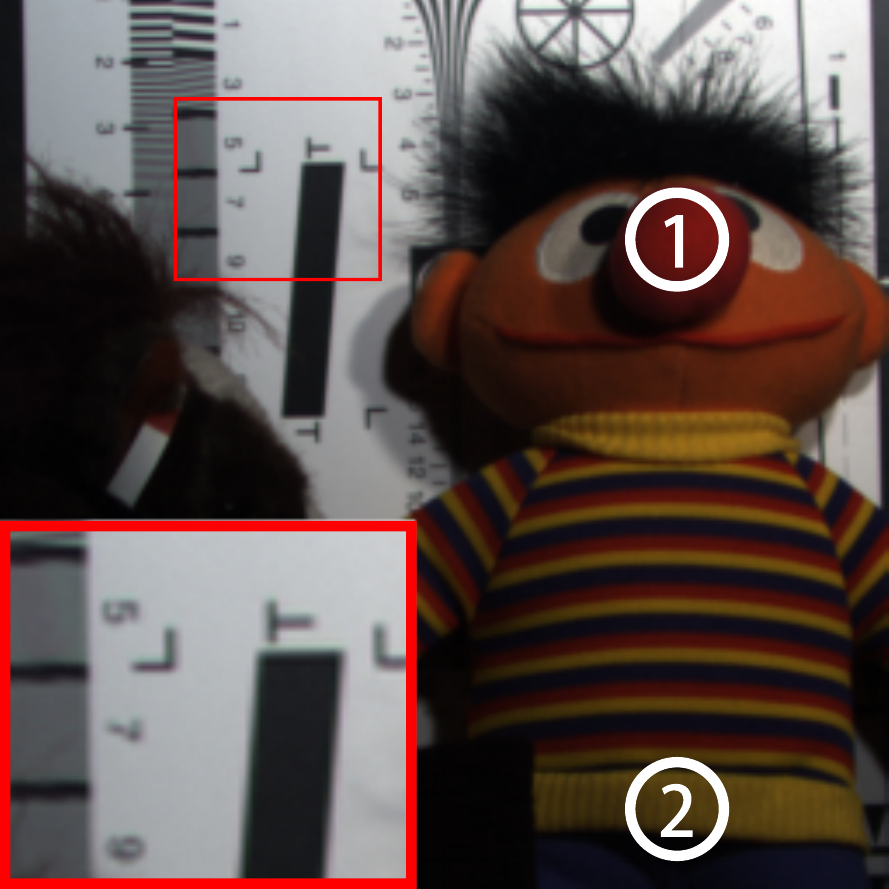}
			%\\  ISTA & & HSCNN  & & HRNet  & & SRP & & Ours  & & Ground Truth
			\\  ISTA & HSCNN & HRNet & SRP &Ours & GT\\
			(18.20 / 0.5173)& 22.66 / 0.8107) & (19.76 / 0.6314) & (22.16 / 0.7421) & (\bf{26.88} / \bf{0.8756})  &  (PSNR  /SSIM)
			\\
			\hline
			\\
			\includegraphics[width=\widthfigure]{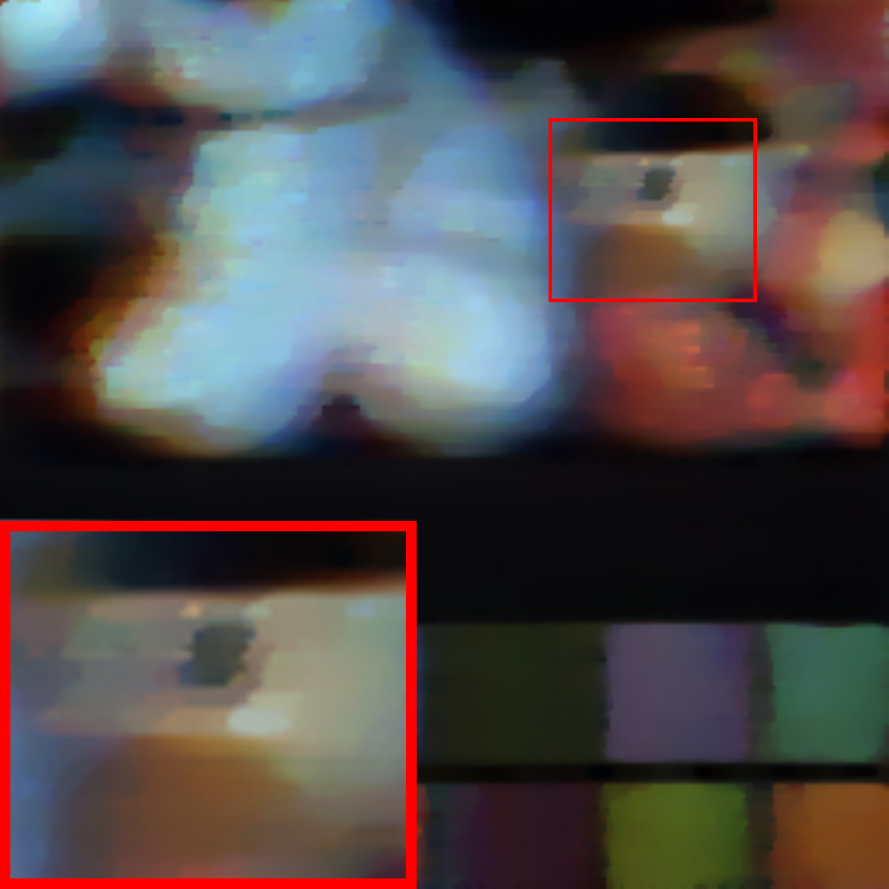}
			&
			\includegraphics[width=\widthfigure]{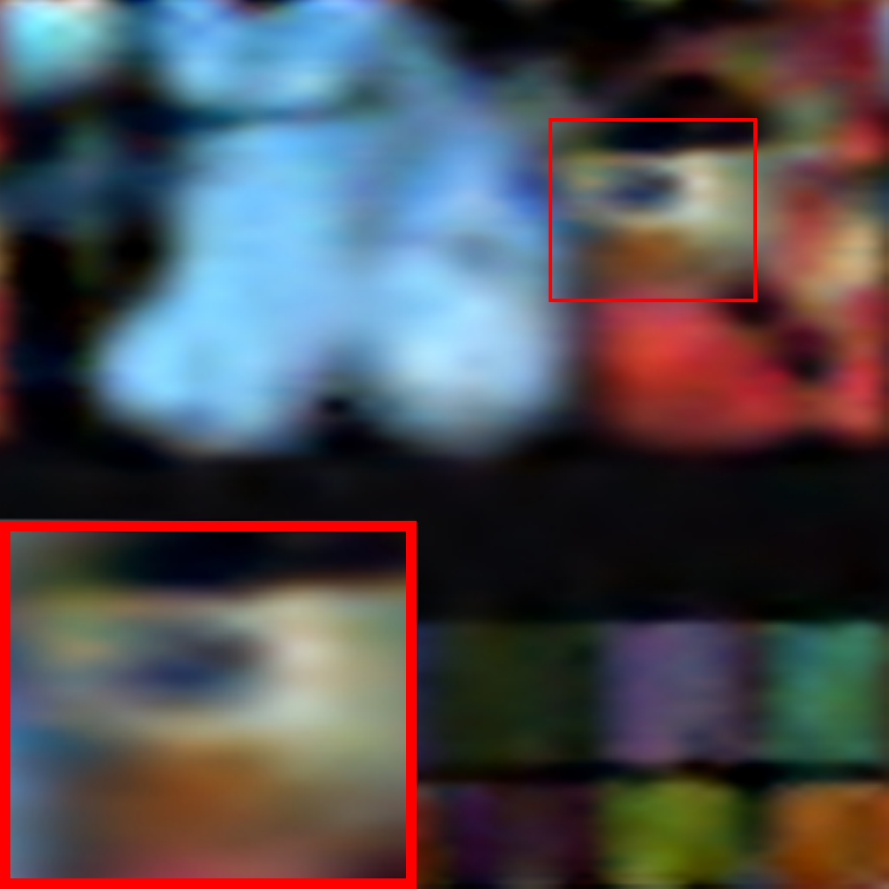}
			&
			\includegraphics[width=\widthfigure]{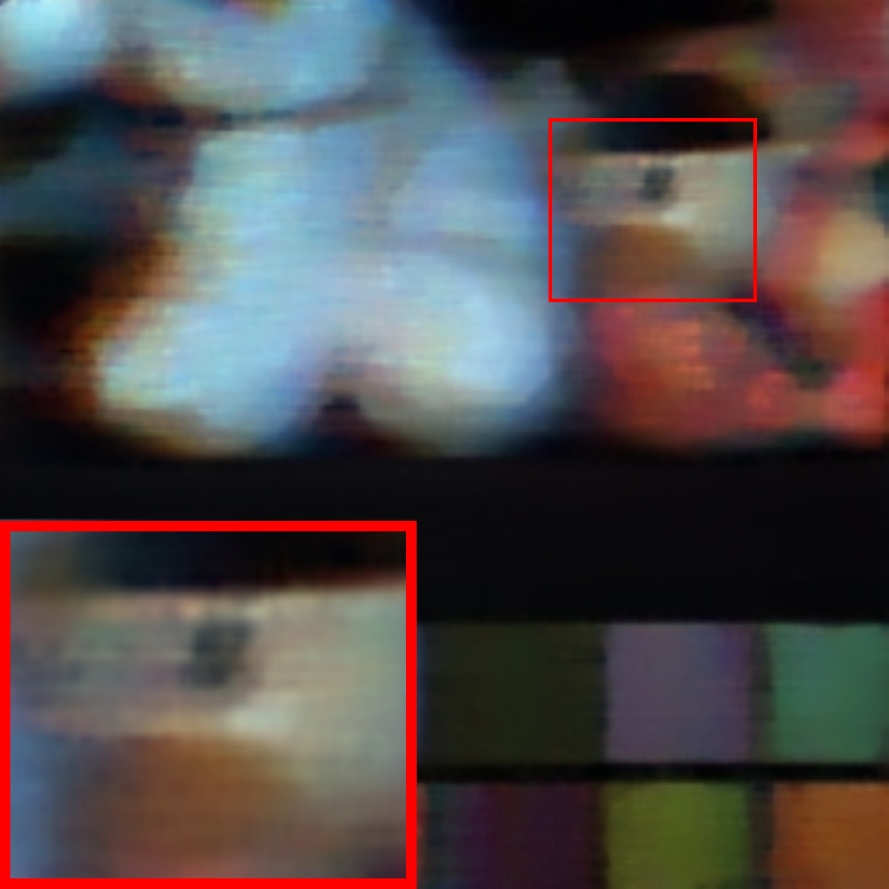}
			&
			\includegraphics[width=\widthfigure]{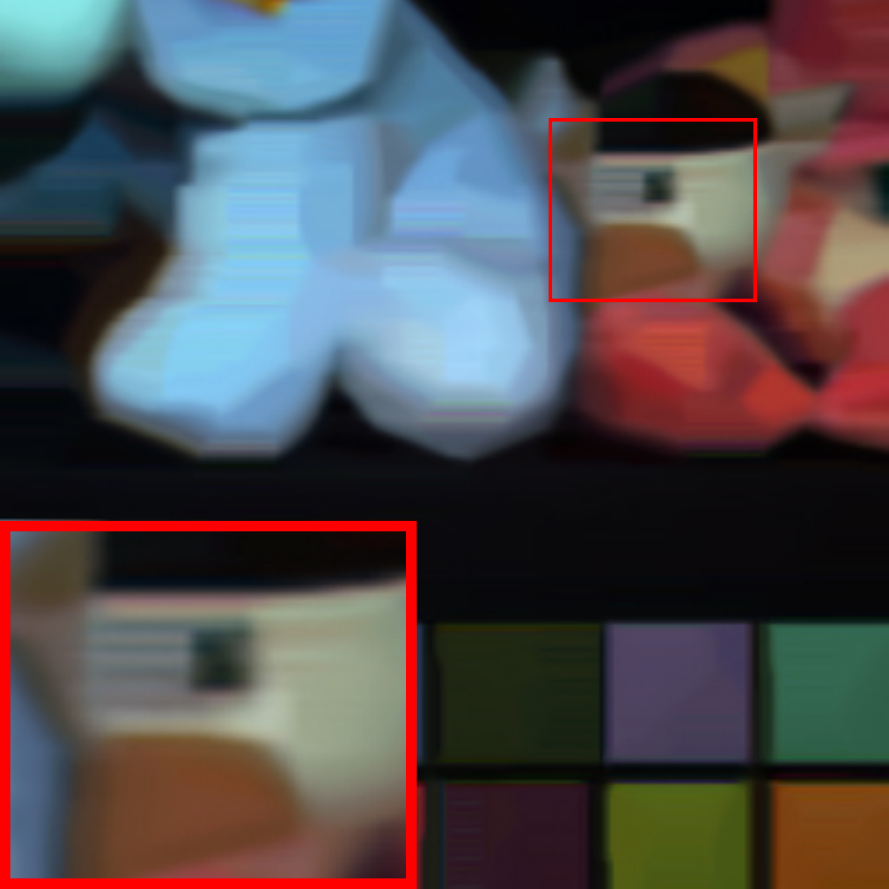}
			&
			\includegraphics[width=\widthfigure]{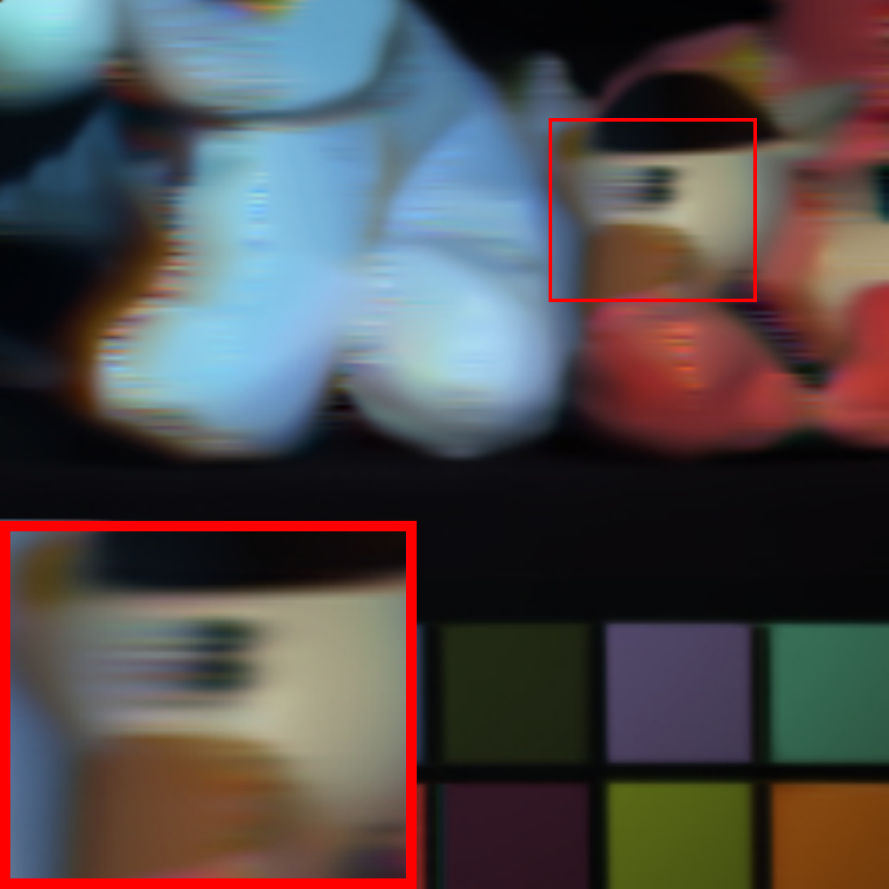}
			&
			\includegraphics[width=\widthfigure]{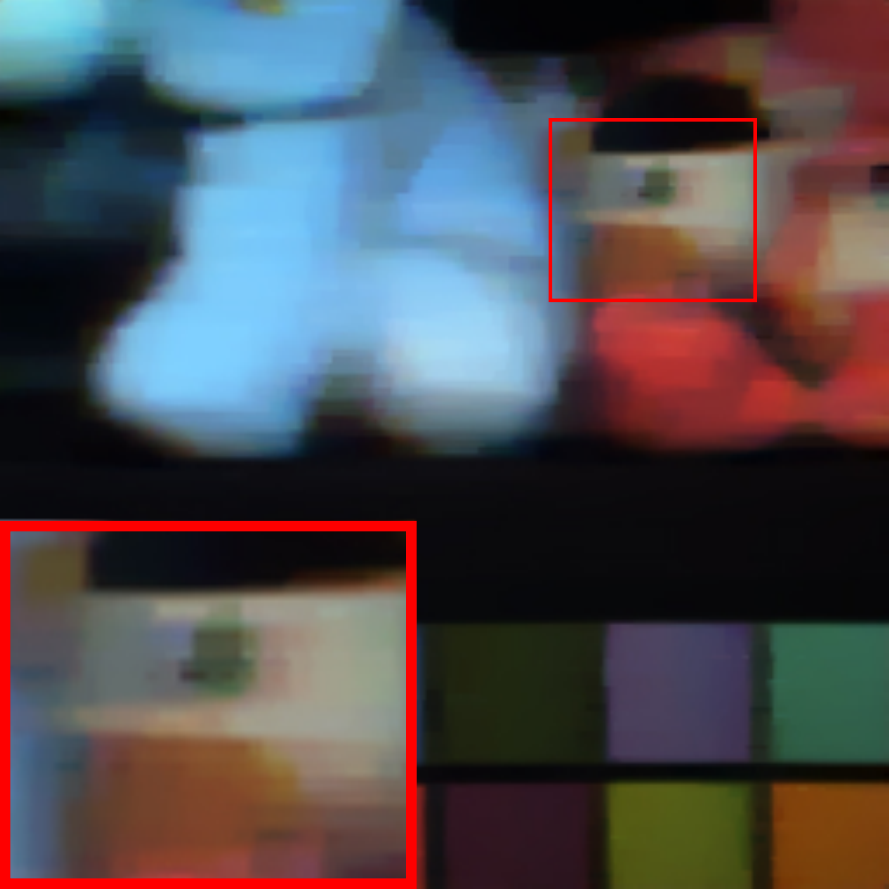}
			\\ TV  & AMP & 3DSR & NSR  & LRMA & AE \\
			(20.42 / 0.7696) & (22.96 / 0.7202) &(22.02 / 0.7796) & (26.90 / 0.8402) & (24.93 / 0.8633)& (26.40 / 0.8496)\\
			\\
			%		\midrule[0.7pt]
			\includegraphics[width=\widthfigure]{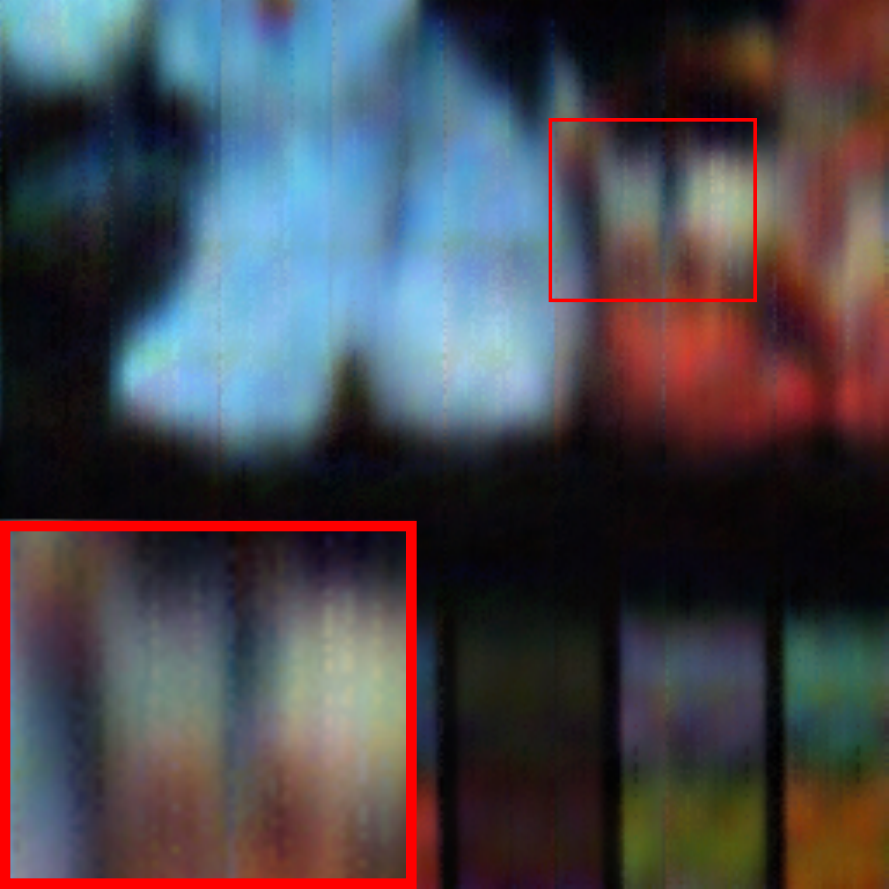}
			&
			\includegraphics[width=\widthfigure]{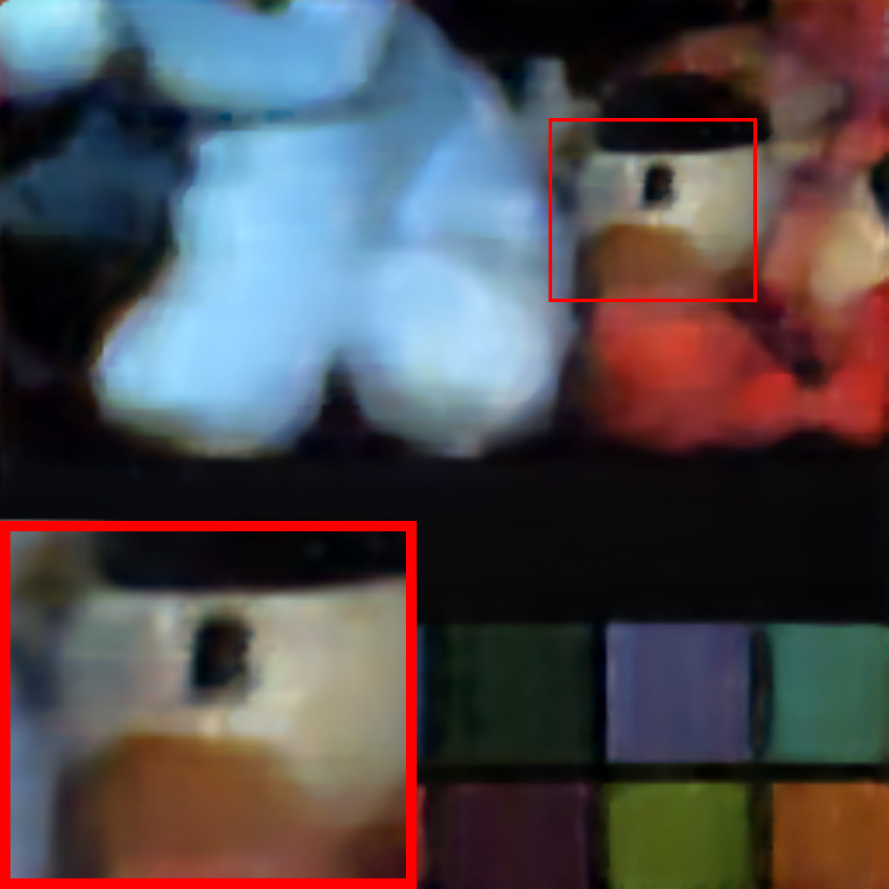}
			&
			\includegraphics[width=\widthfigure]{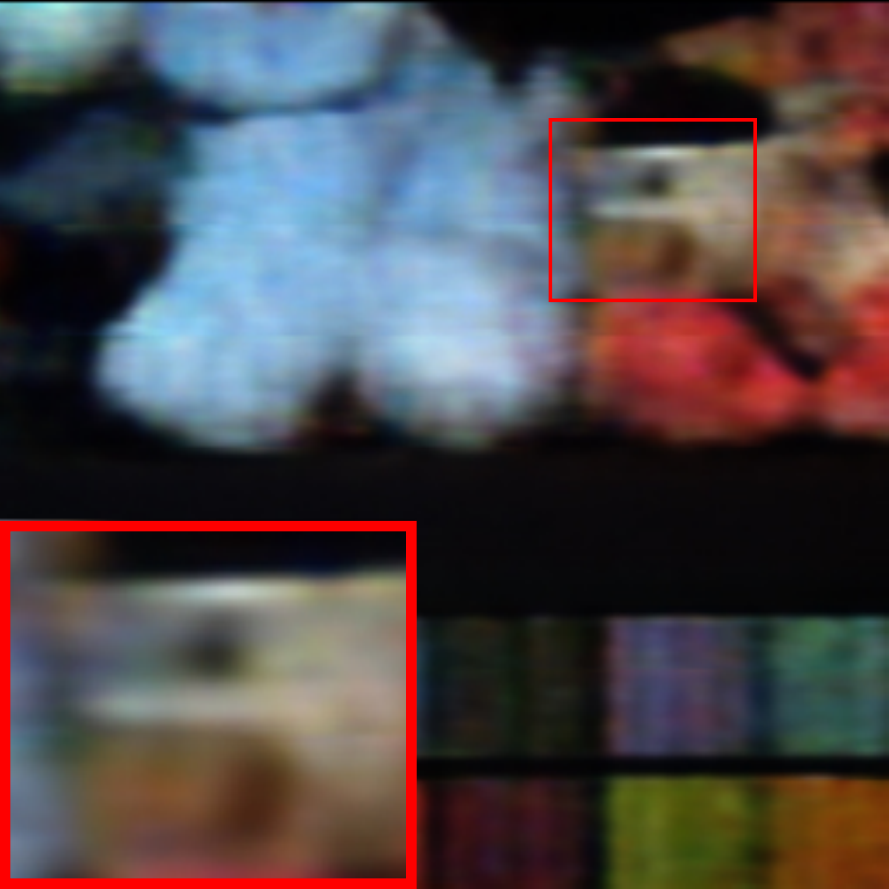}
			&
			\includegraphics[width=\widthfigure]{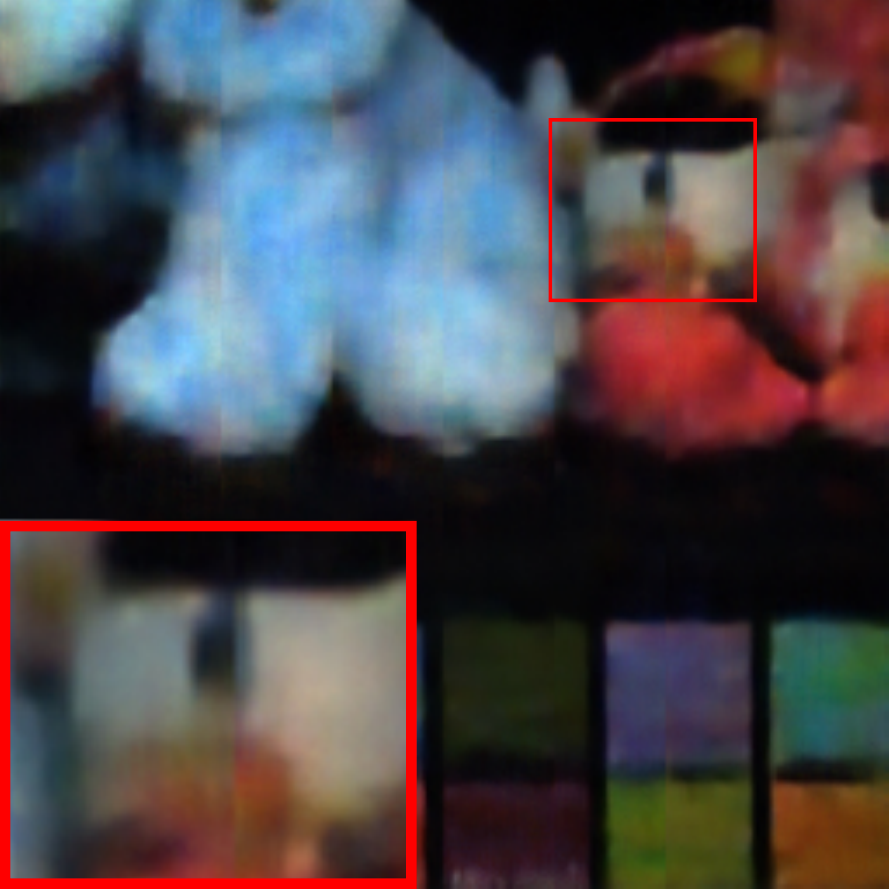}
			&
			\includegraphics[width=\widthfigure]{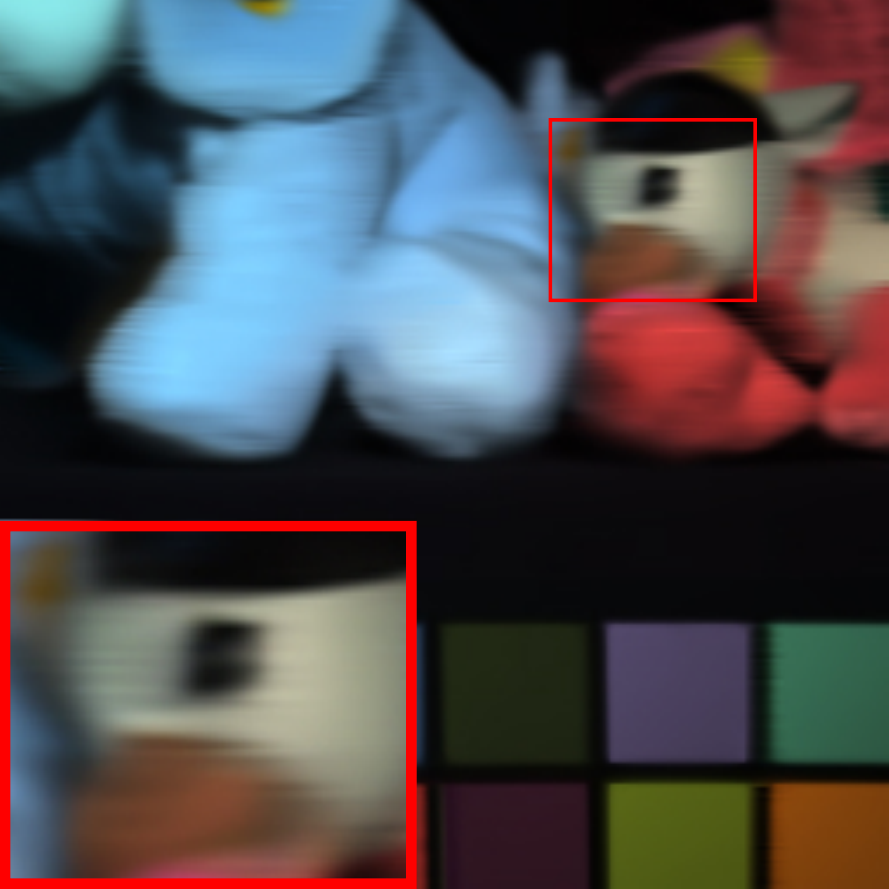}
			&
			\includegraphics[width=\widthfigure]{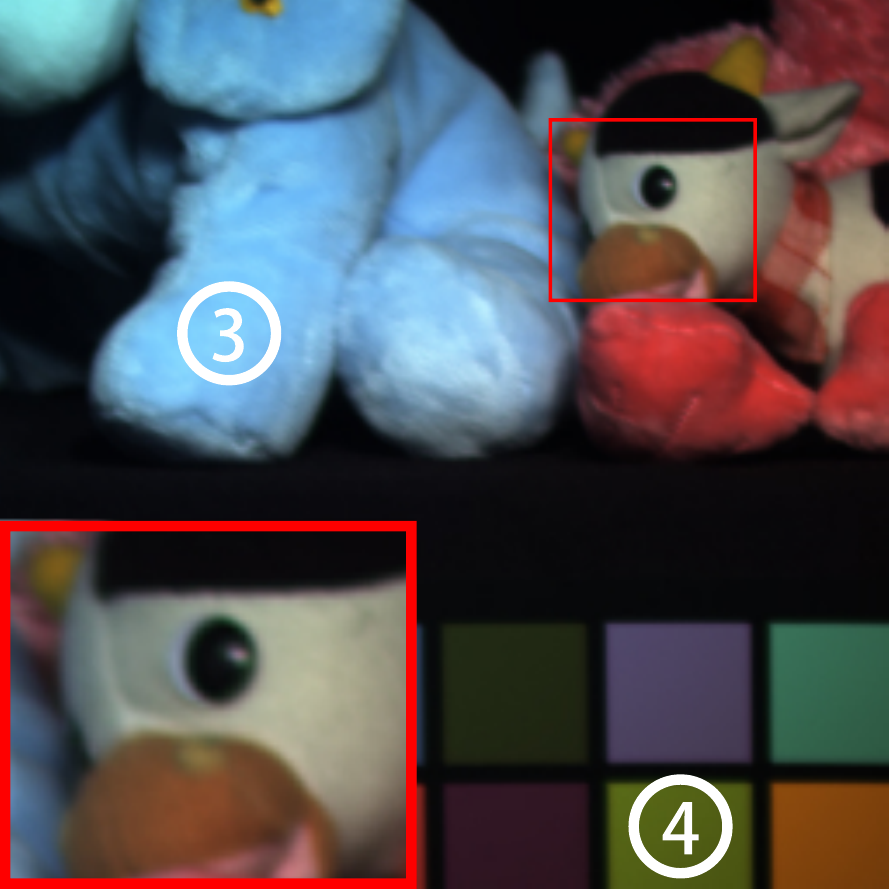}
			%\\  ISTA & & HSCNN  & & HRNet  & & SRP & & Ours  & & Ground Truth
			\\  ISTA &HSCNN & HRNet & SRP &Ours & GT
			\\(19.66 / 0.5135) &(25.80 / 0.8335)& (21.51 / 0.6998) & (23.83 / 0.7675) &(\textbf{30.39} / \textbf{0.9159}) & (PSNR / SSIM)
		\end{tabular}		
		\caption{Reconstructed quality comparison of CASSI. The PSNR and SSIM for the result images of \emph{chart and stuffed toy} and \emph{stuffed toys} are shown in the parenthesis. By comparing the reconstructed results and ground truth (shortened as GT in the figure), our method obtains better spatial contents and textures.}  
		\label{fig:SynCASSI}
	\end{figure*}
}

\def\SynErrorMap{
	\renewcommand\arraystretch{1.3}
	\begin{figure*}[t] 
		\newcommand{\widthfigure}{0.16\linewidth}
		\newcommand{\nullspace}{0.01pt}
		\centering
		\setlength\tabcolsep{1pt}
		\begin{tabular}{cccccc}
			TV & AMP &3DSR & NSR &LRMA& Ours
			\\
			\includegraphics[width=\widthfigure]{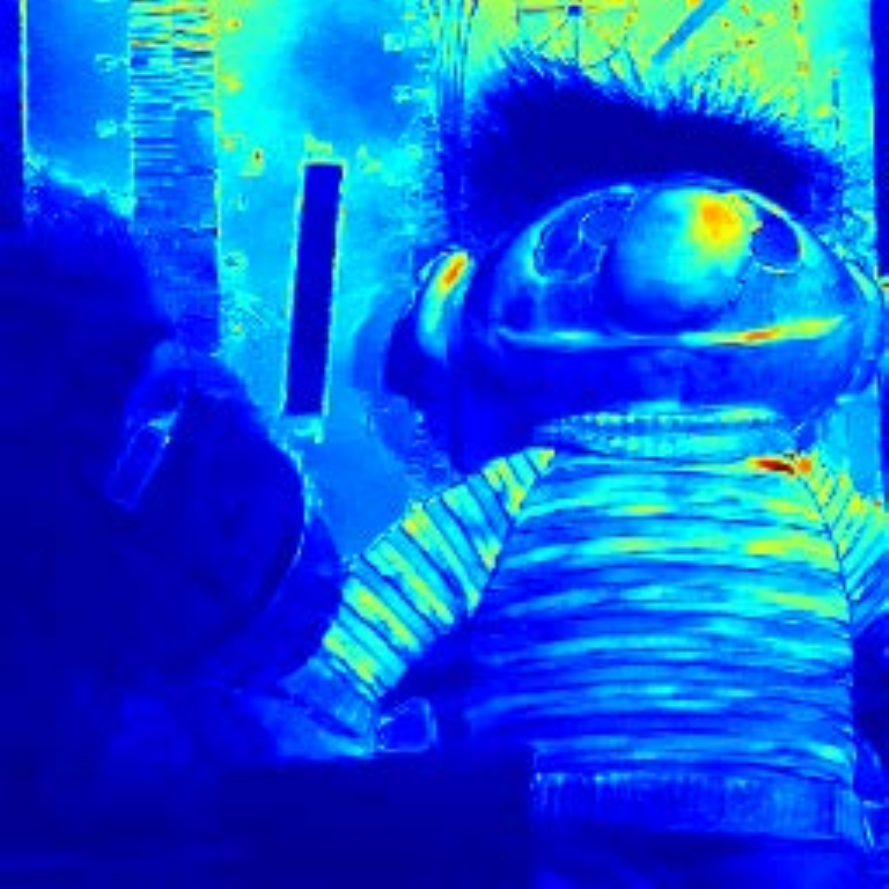}
			&
			\includegraphics[width=\widthfigure]{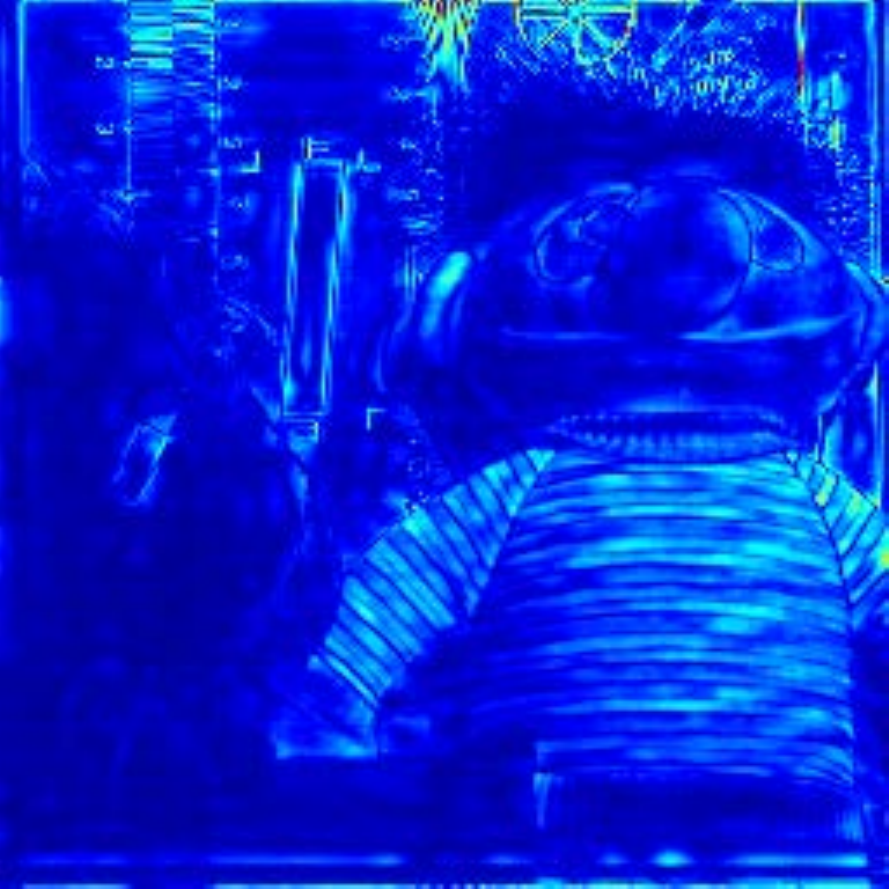}
			&
			\includegraphics[width=\widthfigure]{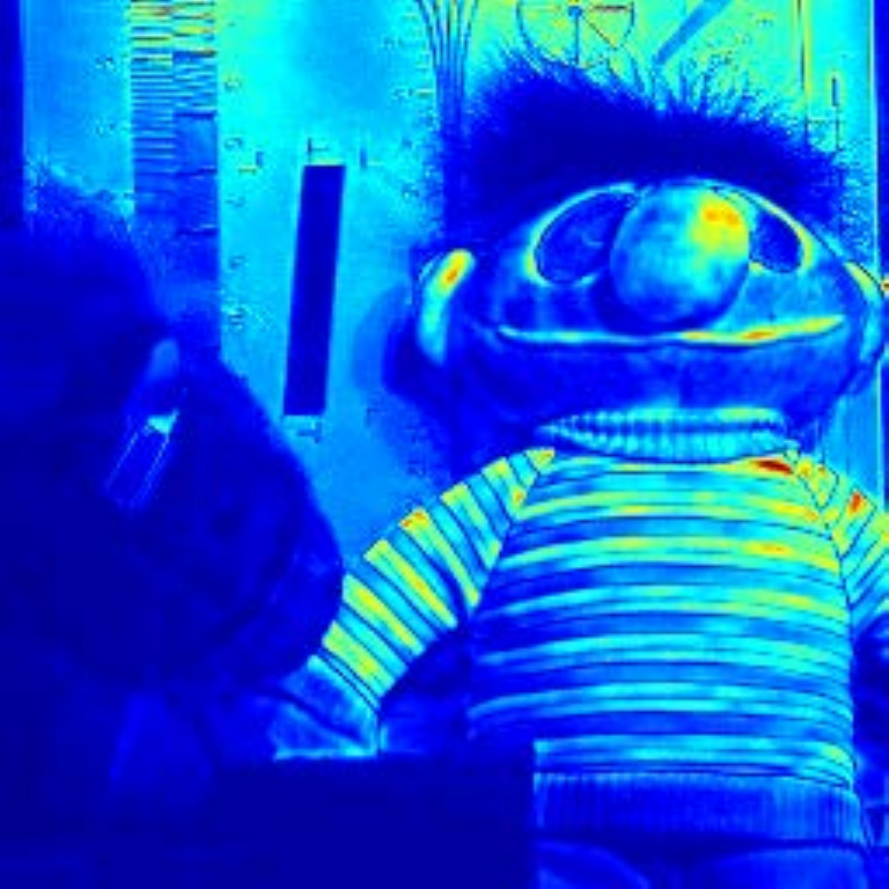}
			&
			\includegraphics[width=\widthfigure]{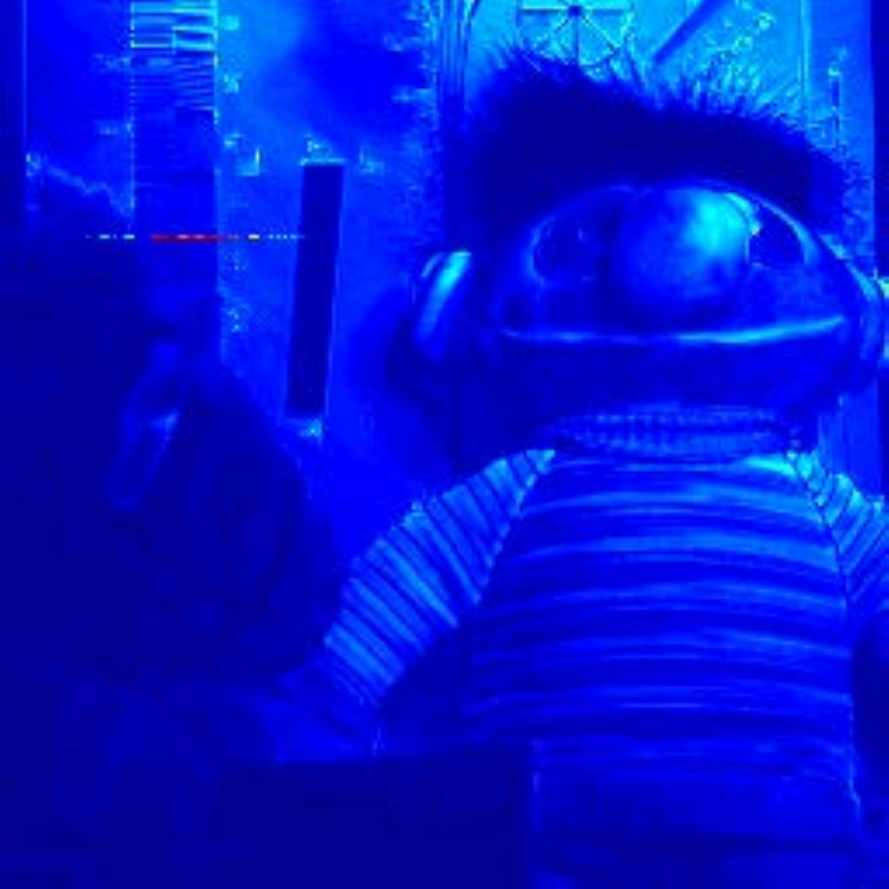}
			&
			\includegraphics[width=\widthfigure]{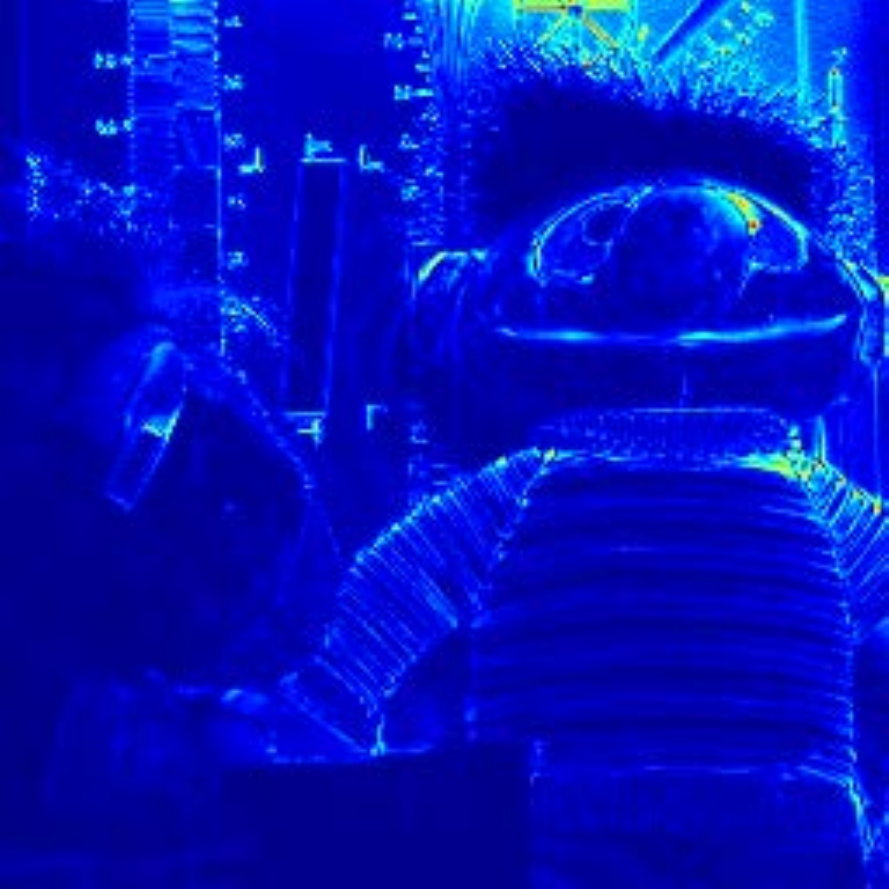}
			&
			\includegraphics[width=\widthfigure]{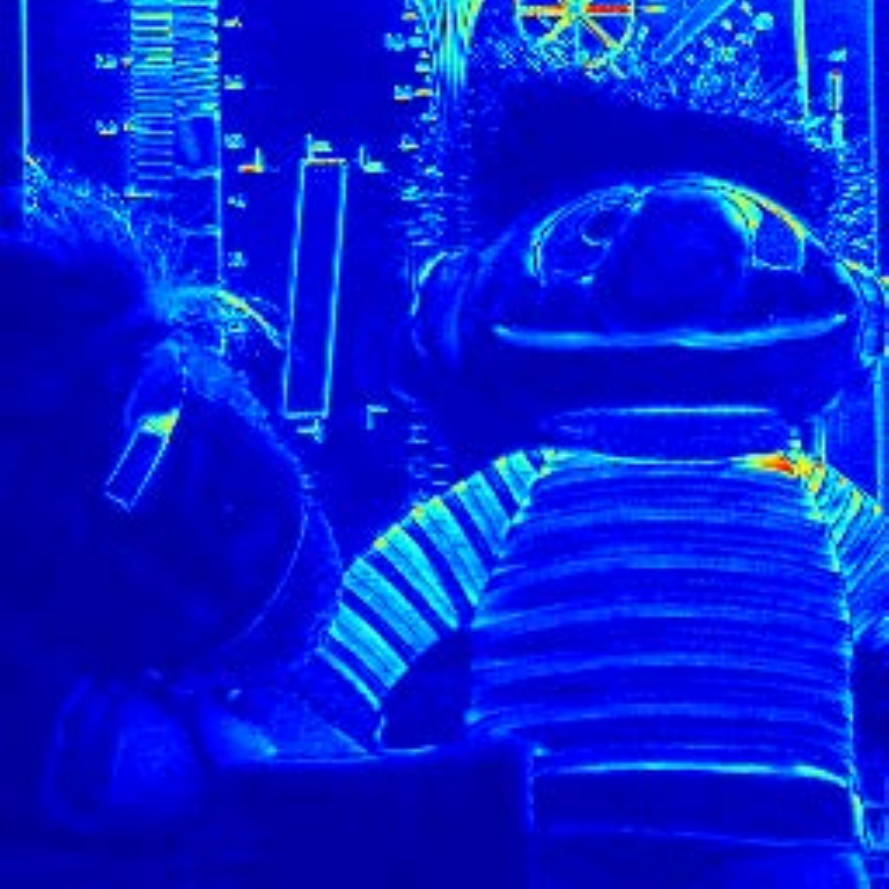}
			\\ 
			%\small{TV} & \small{AMP} &\small{3DSR} & \small{ANSR} & \small{LRMA}& \small{Ours}\\
			%		\midrule[0.7pt]
			\includegraphics[width=\widthfigure]{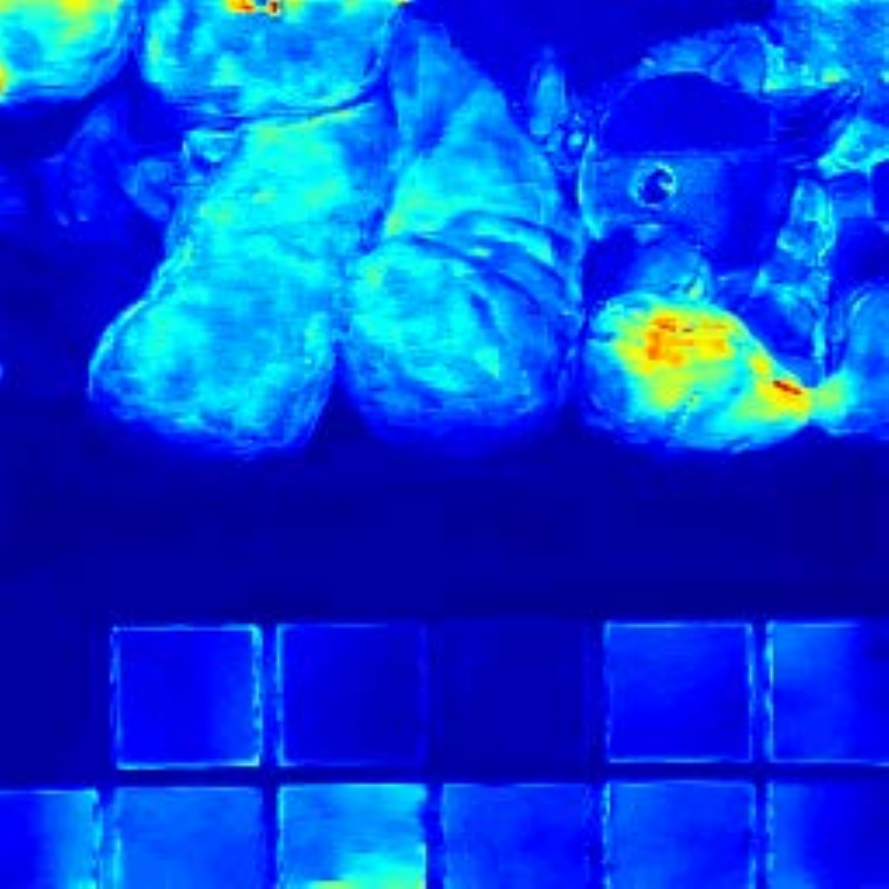}
			&
			\includegraphics[width=\widthfigure]{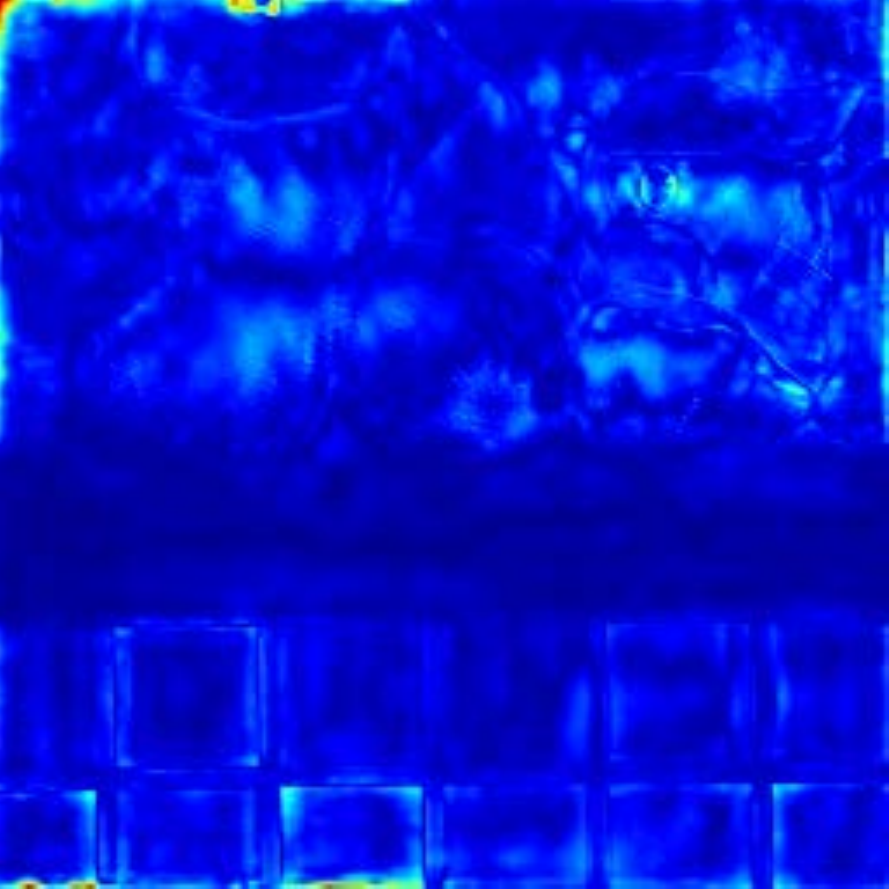}
			&
			\includegraphics[width=\widthfigure]{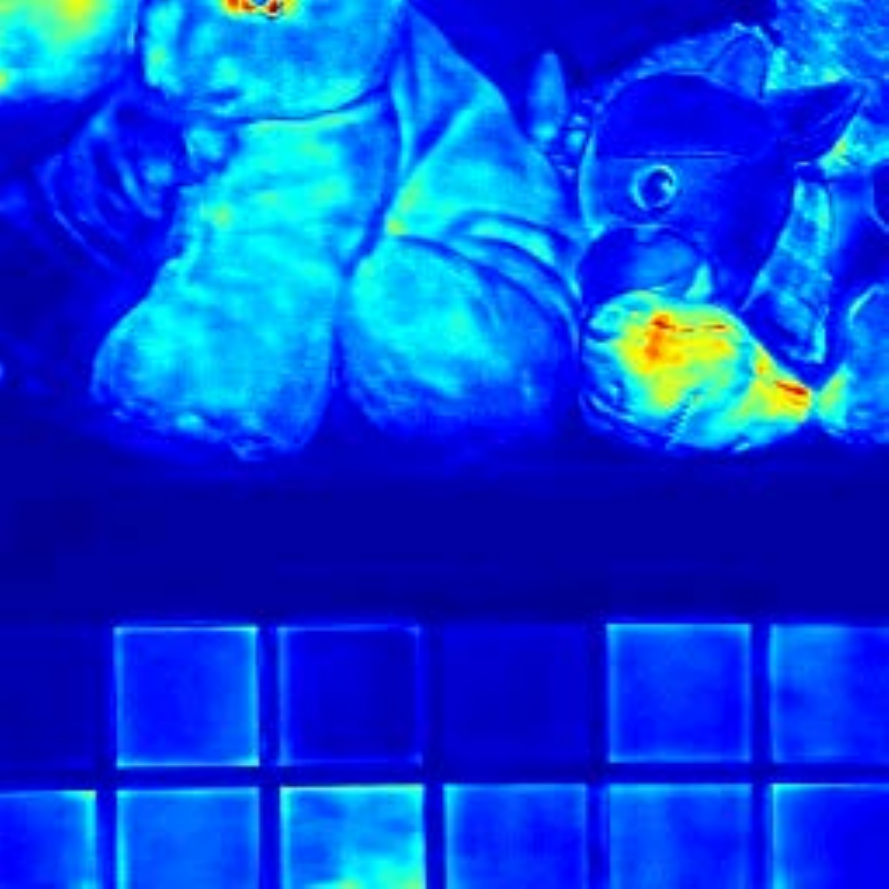}
			&
			\includegraphics[width=\widthfigure]{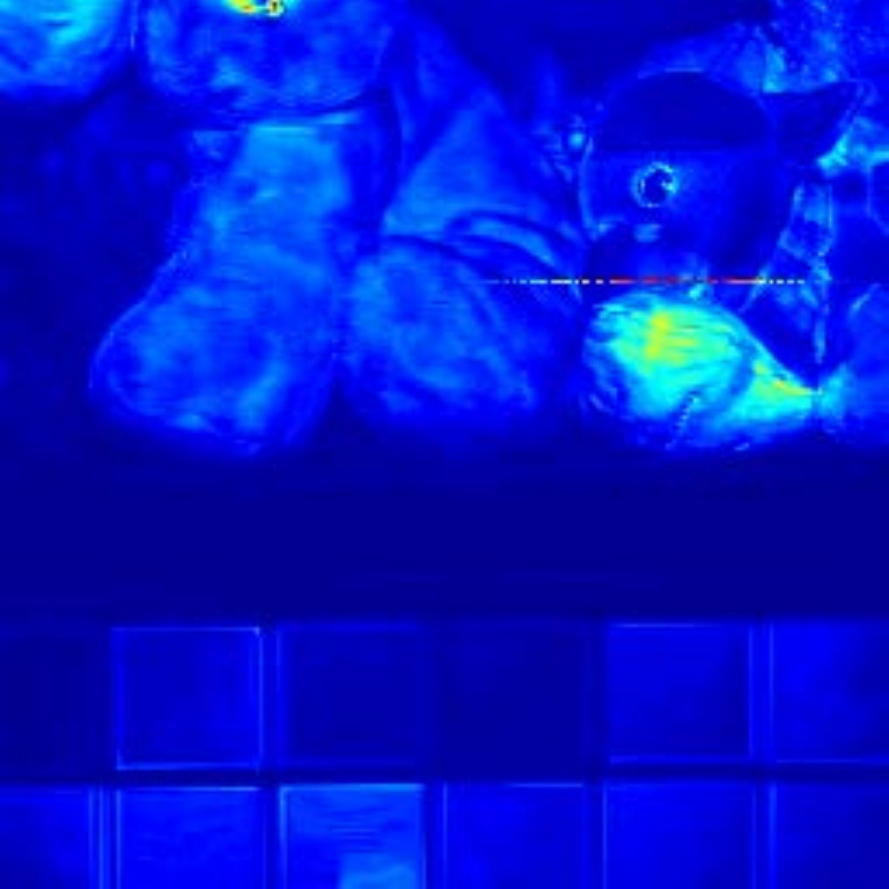}
			&
			\includegraphics[width=\widthfigure]{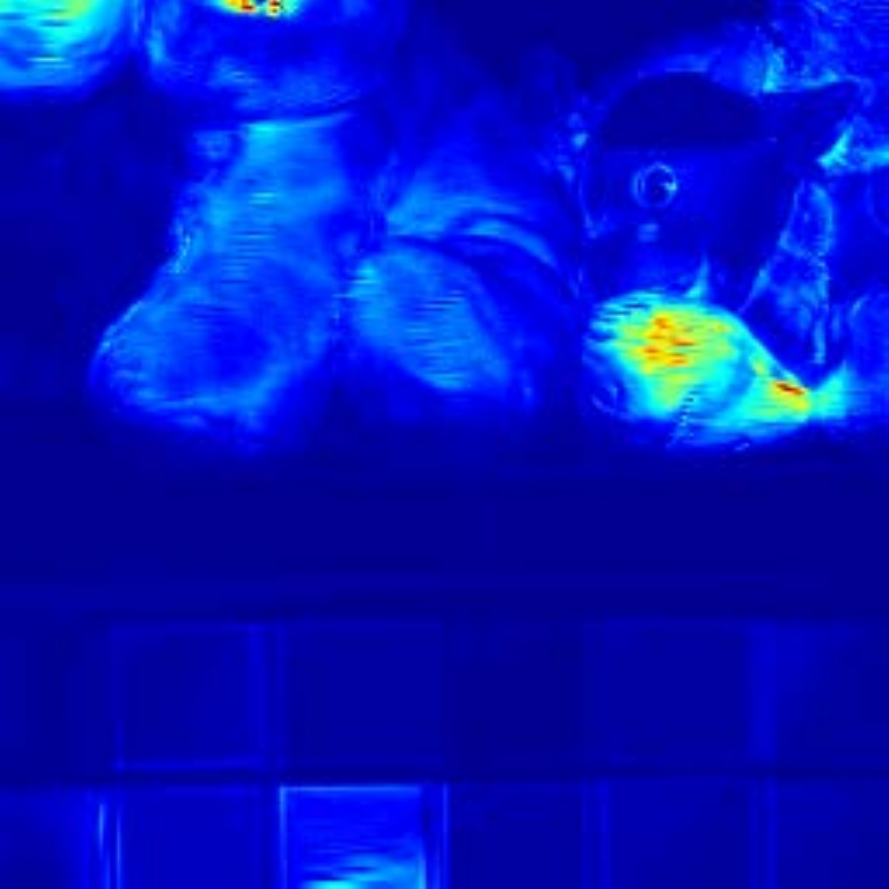}
			&
			\includegraphics[width=\widthfigure]{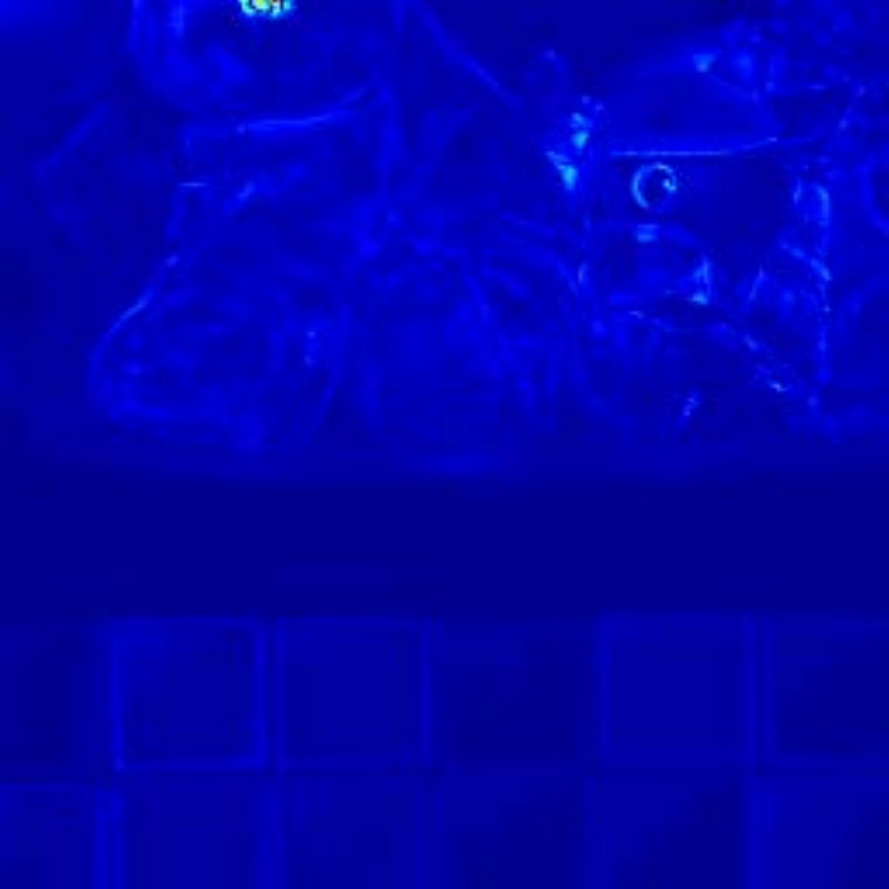}
			%\\  ISTA & & HSCNN  & & HRNet  & & SRP & & Ours  & & Ground Truth
			\\ 
			\multicolumn{6}{c}{\includegraphics[width=13cm]{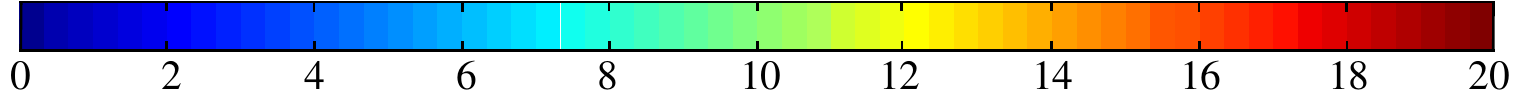}} 
		\end{tabular}		
		\caption{ The error maps comparison of two typical scenes on DCCHI. It shows that our method can produce relatively higher spatial fidelity.}  
		\label{fig:ErrorMap}
	\end{figure*}
}

\def\SynSpectra{
	\renewcommand\arraystretch{1.3}
	\begin{figure*}[t] 
		\newcommand{\widthfigure}{0.240\linewidth}
		\newcommand{\nullspace}{0.01pt}
		\centering
		\setlength\tabcolsep{1pt}
		\begin{tabular}{cccc}
			\includegraphics[width=\widthfigure]{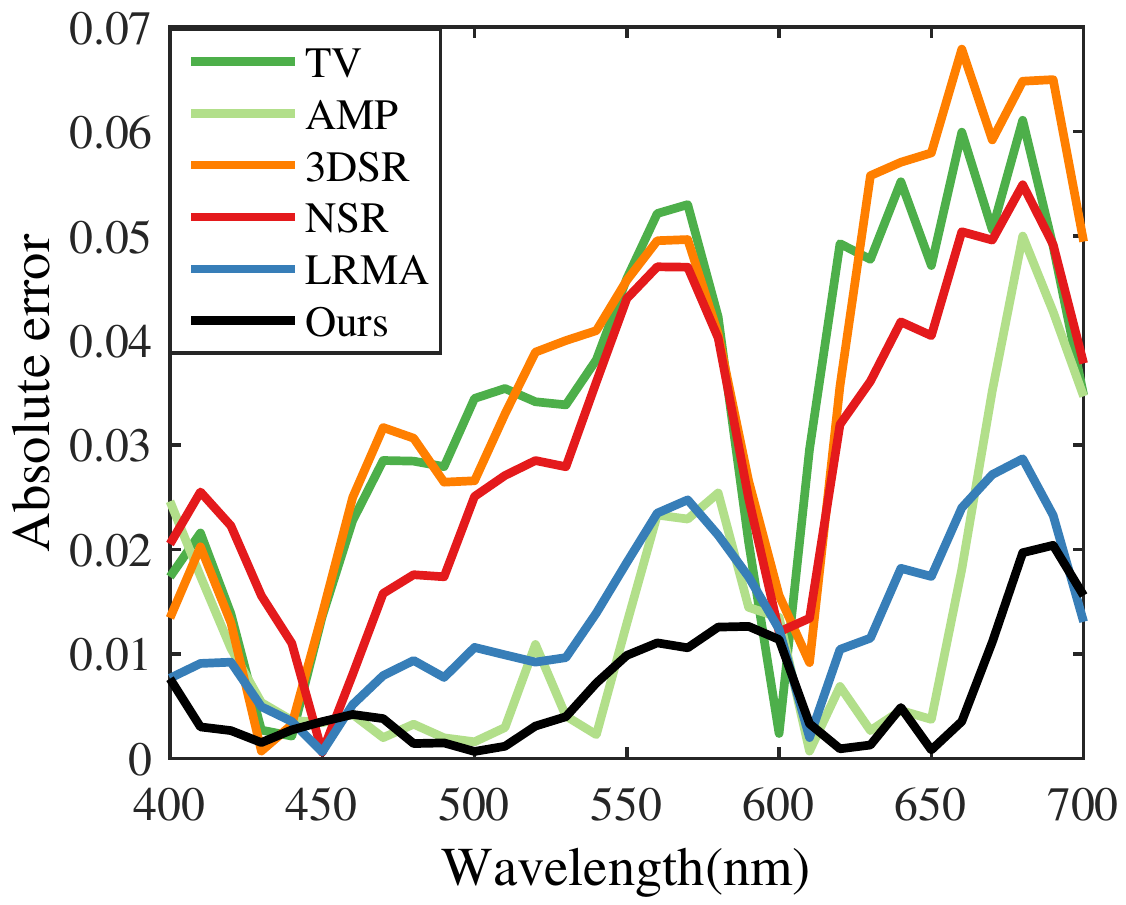}
			&
			\includegraphics[width=\widthfigure]{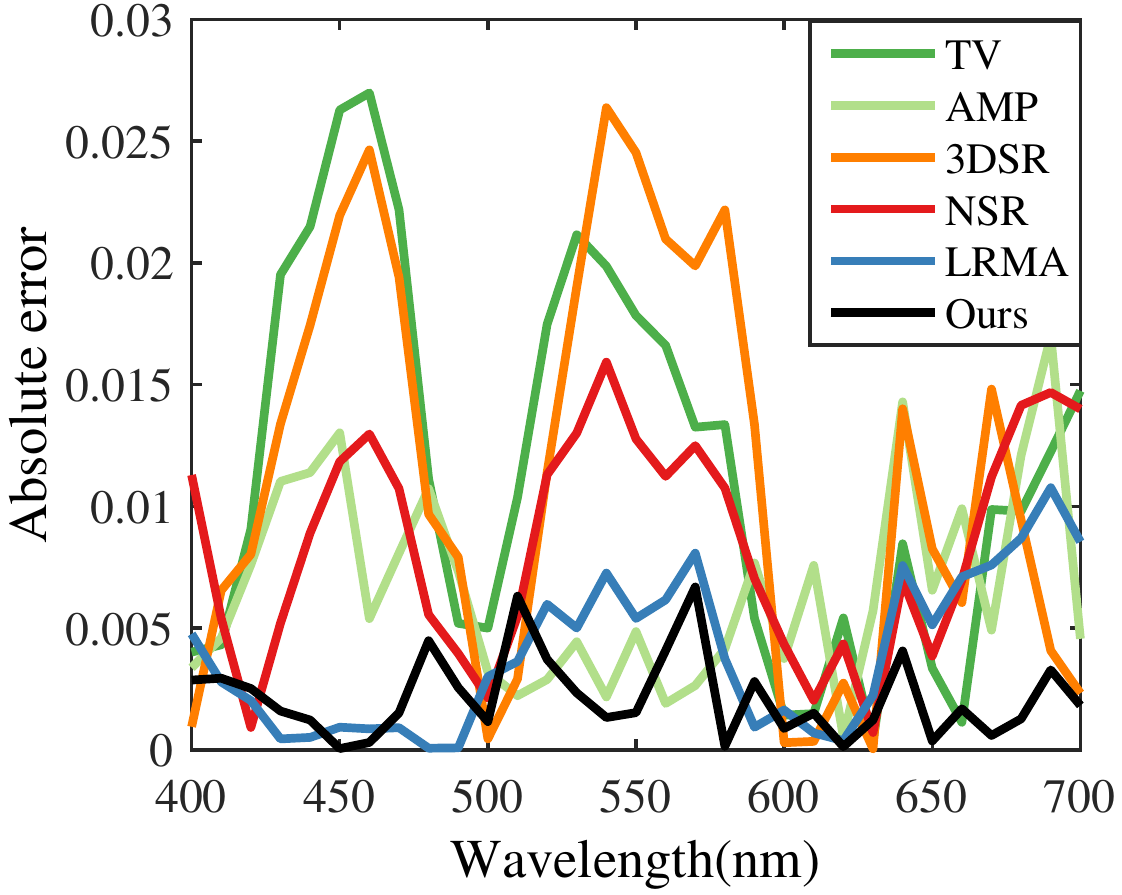}
			&
			\includegraphics[width=\widthfigure]{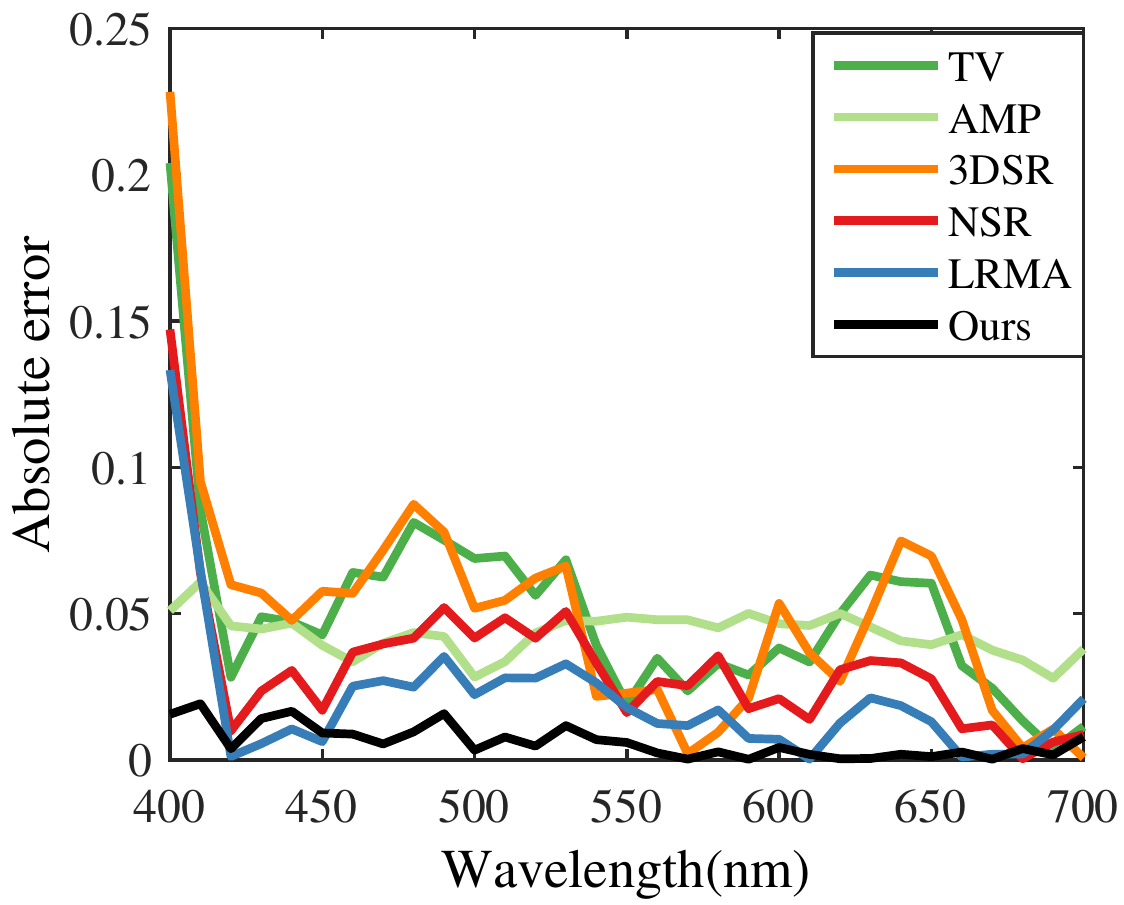}
			&
			\includegraphics[width=\widthfigure]{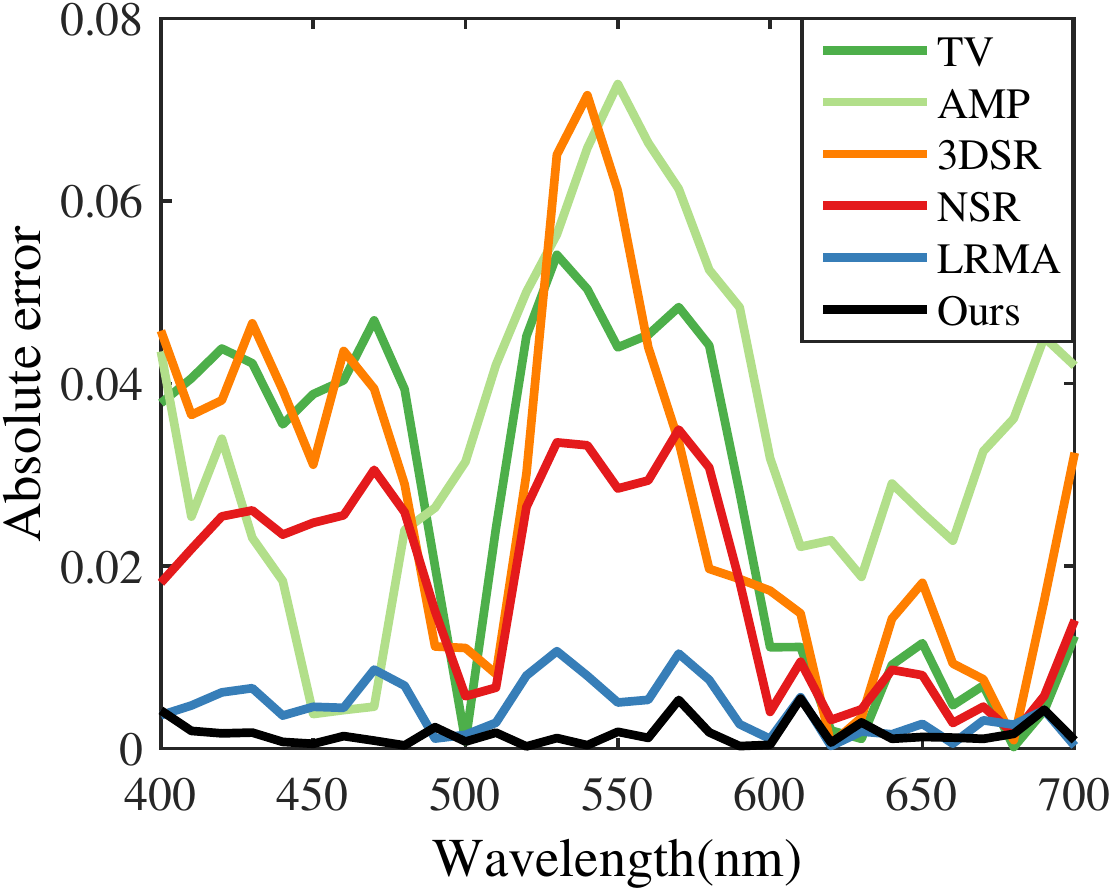}
		\end{tabular}	
		%	\vspace{-0.3cm}	
		\caption{ The absolute spectral error on DCCHI between the ground truth and reconstructed results of the white labels in Fig.~\ref{fig:SynCASSI}. It shows that our method obtains the highest spectral accuracy.}  
		\label{fig:SynSpectra}
		%	\vspace{-0.8cm}
	\end{figure*}
}

\begin{abstract}
Snapshot hyperspectral imaging can capture the 3D hyperspectral image (HSI) with a single 2D measurement and has attracted increasing attention recently. Recovering the underlying HSI from the compressive measurement is an ill-posed problem and exploiting the image prior is essential for solving this ill-posed problem. However, existing reconstruction methods always start from modeling image prior with the 1D vector or 2D matrix and
cannot fully exploit the structurally spectral-spatial nature in 3D HSI, thus leading to a poor fidelity. In this paper, we propose an effective high-order tensor optimization based method to boost the reconstruction fidelity for snapshot hyperspectral imaging. We first build high-order tensors by exploiting the spatial-spectral correlation in HSI. Then, we propose a weight high-order singular value regularization (WHOSVR) based low-rank tensor recovery model to characterize the structure prior of HSI. By integrating the structure prior in WHOSVR with the
system imaging process, we develop an optimization framework for HSI reconstruction, which is finally solved via the
alternating minimization algorithm. Extensive experiments
implemented on two representative systems demonstrate
that our method outperforms state-of-the-art methods.
\end{abstract}

\IEEEpeerreviewmaketitle

\section{Introduction}
Hyperspectral imaging techniques, which are able to capture the 3D hyperspetral image (HSI) with multiple discrete bands at specific frequencies, have attracted increasing interests in recent years. By providing abundant spatial and spectral information simultaneously, HSI can be used for many visual tasks that traditional gray image or color image cannot accomplish. In recent years, HSI has been applied in various vision tasks, such as classification~\cite{Min20123D}, segmentation~\cite{tarabalka2009segmentation}, recognition~\cite{Pan2003Face} and tracking~\cite{Nguyen2010Tracking}.

HSI consists of 2D spatial information and 1D spectral information. To obtain the 3D HSI, conventional hyperspectral imaging techniques scan the scene along 1D or 2D coordinate~\cite{porter1987system,basedow1995hydice,yamaguchi2006high,schechner2002generalized}. This imaging is time-consuming and can not be used in dynamic scenes. To address the limitations of conventional spectral imaging techniques, many snapshot hyperspectral imaging systems based on compressive sensing theory~\cite{Baraniuk2007Compressive,Cand2006Compressive} has been developed. Among those numerous imaging systems, the coded aperture snapshot spectral imager (CASSI)~\cite{Arce2014Compressive,Wagadarikar2008Single},  stands out in the dynamic scene due to its advantage in collecting the 3D data from different wavelengths with one shot. As an advancement system of CASSI, the latest proposed design of dual-camera compressive hyperspectral imaging (DCCHI) incorporates a common panchromatic camera to collect more information simultaneously with CASSI~\cite{Wang2015Dual,Wang2015High}. With the complementary details from the uncoded panchromatic branch, DCCHI can improve the reconstruction fidelity significantly, which obtains the potential ability to apply snapshot hyperspectral imaging into practice~\cite{zhang2018fast}.

Since the number of snapshot hyperspectral imaging systems is far less than what is required by the Nyquist sampling frequency, the reconstruction is severely under-determined. Such a difficult issue can be addressed by solving a convex optimization problem regularized with a HSI prior.
So far considerable reconstruction methods have been proposed in this domain~\cite{Figueiredo2007Gradient,7328255,Lin2014Spatial,7676344,fu2016exploiting,liu2018rank}. However, most of
them start from expressing HSI as 1D vector or 2D matrix and
cannot fully exploit the structurally spectral-spatial prior in 3D HSI. Such a compromise treatment results in poor reconstruction quality and hinder the application of snapshot hyperspectral imaging.

\FigFrame
To overcome the aforementioned drawbacks, we propose a novel tensor based reconstruction method with weight high-order singular value regularization (WHOSVR). Our key observation is that, compared with the 1D vector and 2D matrix, the high-order tensor is more faithful to delivery the structure information of the 3D HSI. Such advantage of tensor motivates us to exploit the high-order representation for snapshot hyperspectral imaging reconstruction and promote the reconstruction fidelity. 
Specifically, we first build a high-order tensor by exploiting the high correlation in the spatial and spectral domains for each exemplar cubic patch. Then, to characterize the high-order structure prior of HSI, we propose a WHOSVR model, where the singular values of tensor are treated adaptively, to finish an accurate low-rank tensor recovery. By integrating the structure prior in WHOSVR with the system
imaging process, we develop an optimization algorithm for
HSI reconstruction, which is finally solved via an alternating minimization algorithm. To the best of our knowledge,
it is the first time to exploiting the prior of HSI with WHOSVR for snapshot hyperspectral imaging. Extensive experiments implemented on both CASSI and DCCHI demonstrate that
our method outperforms state-of-the-art methods.

\section{Related Works}

\subsection{Hyperspectral Imaging System}
Conventional scanning-based hyperspectral imaging systems
directly trade the temporal resolution for the spatial/
spectral
resolution and thus lose the ability to record dynamic scenes~\cite{porter1987system,basedow1995hydice,yamaguchi2006high,schechner2002generalized}.
To capture the dynamic scenes, several snapshot hyperspectral imaging systems have been developed in the last decades~\cite{Cao2011A,cao2016computational, gao2016review}. However, these systems still suffer from the trade-off between the spatial and temporal resolution.

Leveraging the compressive sensing theory, CASSI stands out as a promising solution for the trade-off between the spatial and temporal resolution recently.
CASSI employs one disperser or two dispersers to capture a
2D encoded image of the target~\cite{Arce2014Compressive,Wagadarikar2008Single}. Then the underlying
HSI can be reconstructed from the compressive measurement by solving an ill-posed problem.
Several hardware advancements of CASSI have been
developed to improve the performance, e.g. multiple snapshot
imaging system~\cite{Kittle2010Multiframe,wu2011development} and spatial-spectral compressive
spectral image~\cite{Lin2014Spatial}. 
The latest dual-camera design, i.e., DCCHI \cite{Wang2015Dual,Wang2015High}, incorporates a co-located panchromatic camera to collect more information simultaneously with CASSI. 
With the complementary details from the uncoded image,
DCCHI obtains a significant improvement on reconstruction
quality and the potential ability to be applied into practice. In this paper, we propose a high-order tensor optimization
based method to boost the reconstruction accuracy for snapshot hyperspectral imaging. Meanwhile, our method is a general approach and can be easily extends different systems.

\subsection{Hyperspectral Image Reconstruction}
Recovering the underlying HSI from the compressive measurement plays an essential role for snapshot hyperspectral imaging. Based on compressive sensing, HSI can be reconstructed by solving optimization problems with prior knowledge based regularization. With the hypothesis that nature images owns piecewise smooth property, TV prior based methods have been widely used in snapshot hyperspectral imaging~\cite{Wagadarikar2008Spectral,8529273,zhang2018gpu}. 
By conducting wavelet transform or over-completed learned dictionary as sparsity basis, various sparse reconstruction methods have been developed~\cite{Wagadarikar2008Single,Wang2015High,Figueiredo2007Gradient,7676344,7328255}. Matrix rank minimization regularization based on spectral-spatial correlation are also adopted~\cite{fu2016exploiting,liu2018rank}. However, all these methods always start their modeling with vector or matrix and  ignore the high-dimensionality nature of HSI, thus leading to poor reconstruction quality.
In recent years, several methods on deep learning, including ADMM-Net~\cite{sun2016deep} and ISTA-Net~\cite{zhang2018ista} have been developed for compressive sensing of nature image. However, the heterogeneity of HSI makes those methods difficult to be extended for snapshot hyperspectral imaging. A convolutional autoencoder (AE) based method was first designed for CASSI to learn a non-linear sparse representation~\cite{choi2017high}. A recent work named HSCNN~\cite{xiong2017hscnn}, which treats the reconstruction task as an image enhancement task, was proposed for CASSI. HRNet utilized two separate networks to explore the
spatial and spectral similarity~\cite{wang2018hyperreconnet}. 
The latest design in~\cite{wang2019hyperspectral} combined the neural network and the optimization framework to learn
the spectral regularization prior (SRP) and achieved suboptimal results.
However, these methods all try to learn
a single image prior information with network and still ignore the high-dimensionality intrinsic structure of HSI.

\section{Approach}
\subsection{ Notations and Preliminaries}
We first introduce notations and preliminaries as follows. Matrices are denoted as boldface capital letters, e.g., $\bm{X}$, vectors are represented with blodface lowercase letters, e.g. $\bm{x}$, and scalars are indicated as lowercase letters, e.g., $x$. A tensor of order $N$ as boldface Euler script ${\bm{\mathcal{X}}} \in {\mathbb{R}^{{I_1} \times {I_2} \times  \cdot  \cdot  \cdot {I_N}}}$. By varying index $i_n$ while keeping the others fixed, the mode-$n$ fiber of \bm{${\mathcal{X}}$} can be obtained. By arranging the mode-$n$ fibers of $\bm{\mathcal{X}}$ as column vectors, we get the mode-$n$ unfolding matrix $\bm{X}_{(n)}\in {\mathbb{R}^{{I_n} \times ({I_1}\cdot  \cdot  \cdot {I_{n-1}}{I_{n+1}}\cdot  \cdot  \cdot {I_N})}}$. $\text{fold}_n(\cdot)$ is the operator that converts the matrix back to the tensor format along the mode-$n$.

The \emph{Frobenius} norm of tensor is defined as the square root of the sum of the squares of all its elements, i.e.,
\begin{equation}
{\left\| {\bm{\mathcal{X}}} \right\|_F} = \sqrt {\sum\nolimits_{{i_1} = 1}^{{I_1}} {\sum\nolimits_{{i_2} = 1}^{{I_2}} { \cdot  \cdot  \cdot \sum\nolimits_{{i_N} = 1}^{{I_N}} {x({{i_1},{i_2} ,\cdot  \cdot  \cdot ,{i_N}})^2} } } }.
\end{equation}

The tensor $n$-mode product is defined as multiplying a tensor by a matrix in mode-$n$. For example, the $n$-mode product of 
${\bm{\mathcal{X}}} \in {\mathbb{R}^{{I_1} \times {I_2} \times  \cdot  \cdot  \cdot {I_N}}}$ by a matrix ${\bm{A}} \in {\mathbb{R}^{{J} \times {I_n}}}$, which denoted as ${\bm{\mathcal{X}}}~{\times}_n~{\bm{A}}$, is an $N$-order tensor ${\bm{\mathcal{B}}} \in {\mathbb{R}^{{I_1} \times\cdot  \cdot  \cdot  {J} \times  \cdot  \cdot  \cdot {I_N}}}$, with entries
\begin{equation}
b({i_1},...,{i_{n - 1}},j,{i_{n - 1}},...,{i_N}) = \sum\limits_{{i_n} = 1}^{{i_N}} {x\left( {{i_1},{i_2},...,{i_N}} \right)a\left( {j,{i_n}} \right)}.
\end{equation}

With the Tucker decomposition \cite{tucker1966some}, an $N$-order tensor ${\bm{\mathcal{X}}}$ can be decomposed into the following form:
\begin{equation}
{\bm{\mathcal{X}}} = \bm{\mathcal{G}}~{\times}_1~ {\bm{U}_1}~{\times}_2~{\bm{U}_2}~{\times}_3\cdot \cdot  \cdot{\times}_N{\bm{U}_N},
\end{equation}
where ${\bm{\mathcal{G}}} \in {\mathbb{R}^{{R_1} \times {R_2} \times  \cdot  \cdot  \cdot {R_N}}}$ is called the core tensor, which is similar to the singular values in matrix SVD, and ${\bm{U}_i} \in {\mathbb{R}^{{I_i} \times {R_i}}}$ is the orthogonal base matrix, which is also similar to the principal components matrix in matrix SVD. Therefore, the Tucker decomposition can be regarded as high-order SVD (HOSVD).
\FigSystem
\subsection{Observation Model}
We then give a brief introduction to the observation model of the two representative systems, i.e., CASSI and DCCHI. It is worth mention that our method is general in this filed and suited for other systems, such as the multiple snapshot imaging system and the spatial-spectral encoded imaging system.

As shown in Fig.~\ref{fig:DualCamera},  
the incident light in CASSI is first projected on the plane of a coded aperture through a objective lens. After spatial coding by the coded aperture, the modulated light goes through a relay lens and is spectrally dispersed in the vertical direction by Amici prism. Finally, the modulated and dispersed spectral information is captured by a panchromatic
camera. Let $\bm{\mathcal{F}}\in {\mathbb{R}^{{I} \times {J} \times {\Lambda}}}$ denote the original HSI and $f({i,j,\lambda})$ is its element, where $1 {\leq} i
{\leq} I$, $1 {\leq} j {\leq} J$ index the spatial coordinate and $ 1 {\leq}
\lambda {\leq} \Lambda$ indexes the spectral coordinate. The compressive measurement at position $(i,j)$ on the focal plane of CASSI can be represented as:
\begin{equation}
%\resizebox{.9\hsize}{!}{$
{y^c}(i,j) = \sum\nolimits_{\lambda  = 1}^\Lambda  {\rho (\lambda )} \varphi(i- \phi (\lambda ),j)f(i - \phi (\lambda ),j,\lambda ),\label{eqFord}
%$}
\end{equation}
where $\varphi(i,j)$ denotes the modulation pattern of the coded aperture, $\phi(\lambda)$ denotes the
dispersion introduced by Amici prism and $\rho(\lambda)$ is
the spectral response of the detector. For brevity, let ${\bm{Y}}^c\in {\mathbb{R}^{{(I + \Lambda-1)} \times J}}$ denote the matrix representation of ${y^c}(i,j)$, and  $\bm{\Phi}^c$ denote the forward imaging function of CASSI, which is jointly determined by $\rho(\lambda)$, $\varphi(i,j)$ and $\phi (\lambda )$. Then the matrix form of CASSI imaging can be expressed as:
\begin{equation}
{\bm{Y}}^c = \bm{\Phi}^c(\bm{\mathcal{F}}).
\label{eqpanFordMtx}
\end{equation}

As shown in Fig.~\ref{fig:DualCamera}, DCCHI consists of a CASSI branch and a panchromatic camera branch. The incident light first is divided into two directions by the beam splitter equivalently. The light in one direction is captured by the CASSI system according to the above imaging principle, while the light on the other direction is captured directly by a panchromatic camera. The compressive measurement ${y^p}(i,j)$ on the panchromatic detector can be represented as:
\begin{equation}
{y^p}(i,j) = \sum\nolimits_{\lambda  = 1}^{\Lambda}  {\rho (\lambda )} f(i,j,\lambda ).
\label{eqpanFord}
\end{equation}

Let ${\bm{Y}}^{p}\in {\mathbb{R}^{{I  \times J}}}$ denote the matrix representation of ${y^p}(i,j)$, and  $\bm{\Phi}^p$ denote the forward imaging function of the panchromatic camera, which is determined by $\rho(\lambda)$. Then the matrix form of the panchromatic branch can be expressed as:
\begin{equation} \label{eqn2}
{\bm{Y}}^p = \bm{\Phi}^p(\bm{\mathcal{F}}),
\end{equation}

A general imaging representation of snapshot hyperspectral imaging can be formulated as:
\begin{equation}\label{eqUniform}
{\bm{Y}} = \bm{\Phi}(\bm{\mathcal{F}}).
\end{equation}
\FigTSVD
For CASSI, $\bm{Y}=\bm{Y}^c$ and $\bm{\Phi}=\bm{\Phi}^c$. 
For DCCHI, $\bm{Y}=[\bm{Y}^c;\bm{Y}^p]$ and $\bm{\Phi}=[\bm{\Phi}^c;\bm{\Phi}^p]$.
The goal of HSI reconstruction is to estimate $\bm{\mathcal{F}}$ from the compressive measurement $\bm{Y}$.

\subsection{Weighted High-order Singular Value Regularization}

The key of reconstruction algorithm is to fully exploit the prior information of HSI and build a suitable regularization model. So far most of existing reconstuction methods are on account of two important properties of HSI, i.e., the spatial self-similarity and the spectral correlation. The spatial self-similarity states a nature that there are many image patches around each exemplar patch with the same texture structure. While the spectral correlation indicates that HSI contains a small amount of basis materials and thus exhibits rich redundancy in spectra. In this paper, we utilize tensor low-rank regularization to take such two properties into consideration simultaneously and promote the reconstruction accuracy.

\TabSynCASSI

For one cubic patch with the size of ${s}\times {s}\times\Lambda$ across full bands of HSI $\bm{\mathcal{F}}\in {\mathbb{R}^{{I} \times {J} \times {\Lambda}}}$, we
search for its $k-1$ nearest neighbors patches in a local
window. By reordering the spatial
block of each band into a 1D column vector, the constructed 3-order tensor $\bm{\mathcal{S}}$ with the size of $s^2 \times \Lambda \times k$ is formed. The constructed tensor simultaneously exhibits the spatial self-similarity (mode-1), the spectral correlation (mode-2) and the joint correlation (mode-3). Here we take the multi-dimensionality property of tensor and introduce low-rank tensor recovery model to preserve the structure information of HSI. Consequently, we build a basic regularization towards low-rank tensor recovery:
\begin{equation}
%{\bm{\mathcal{X}}} = \mathop {\arg \min }
{\Gamma(\bm{\mathcal{S}})} = \tau{  \left\| {{{\bf{R}}(\bm{\mathcal{F}})} - \bm{\mathcal{S}}} \right\|_F^2}+ \text{rank}(\bm{\mathcal{S}}),
\label{eqTensorApro}
\end{equation}
where ${\bf{R}}(\bm{\mathcal{F}})$ represents extracting the 3D tensor from $\bm{\mathcal{F}}$ and $\tau$ denotes the penalty factor.
With the Tucker decomposition \cite{tucker1966some}, the rank of a 3-order tensor can be defined as the sum of ranks of unfolding matrices along three modes. However, considering the rank of different modes separately still ignore the correlation between different modes. Meanwhile, estimating the rank of a matrix is still a NP-hard problem. 

We return to the definition of Tucker decomposition and analyze the low rank property of the constructed tensor. As shown in Fig.~\ref{fig:SVD}, we implement Tucker decomposition on a nonlocal tensor extracted from a clean HSI and show the distribution of singular values in the core tensor. We can see that the singular values tend
to be dropping to zero fleetly, which indicates that the core tensor exhibits significant sparsity.
%Meanwhile, inspired by the success of weighted neuclear norm minimization for low-rank matrix recovery, we introduce the WHOSVR 
Therefore, we can introduce a weighted $\ell_{1}$ norm
to pursue the tensor rank minimization, i.e.,:
\begin{equation}
\text{rank}(\bm{\mathcal{S}}) = \left\| \bf{w} \circ \bm{\mathcal{G}} \right\|_1~~s.t. \bm{\mathcal{S}} = \bm{\mathcal{G}}~{\times}_1~ {\bm{U}_1}~{\times}_2~{\bm{U}_2}~{\times}_3~{\bm{U}_3},
\label{eqWHOSVR1}
\end{equation}
where $\left\| \bf{w} \circ \bm{\mathcal{G}} \right\|_1 = \sum\nolimits_n {{w_n}\left| {{g_n}} \right|} $ and ${g_n}$ is the element of $\bm{\mathcal{G}}$. The weight $w_n$ is set as:
\begin{equation}
w_n^{t+1} = {c \mathord{\left/
		{\vphantom {1 {\left( {\left| {w_i^t} \right| + \varepsilon } \right)}}} \right.
		\kern-\nulldelimiterspace} {\left( {\left| {w_i^t} \right| + \varepsilon } \right)}},
\label{eqWeight}
\end{equation}
where $t$ denotes the $t$-th iteration, $c$ is a positive constant number and $\varepsilon \leq 10^{-6}$. Such formulation derives a meaningful outcome that the singular values can be penalized adaptively and structure information can be better preserved. Specifically, those greater singular values in the $t$-th iteration, which deliver
more important structure information, will get a smaller weight and be shrunk less at $(t+1)$-th iteration. 

\SynCASSI

By combining Equation \ref{eqTensorApro} and \ref{eqWHOSVR1}, we obtain the WHOSVR model as:
\begin{equation}
%{\bm{\mathcal{X}}} = \mathop {\arg \min }
{\Gamma(\bm{\mathcal{G}})} = \tau{  \left\| {{{\bf{R}}(\bm{\mathcal{F}})} - \bm{\mathcal{G}}~{\times}_1~ {\bm{U}_1}~{\times}_2~{\bm{U}_2}~{\times}_3~{\bm{U}_3}} \right\|_F^2}+ \left\| \bf{w} \circ \bm{\mathcal{G}} \right\|_1.
\label{eqWHOSVR}
\end{equation}

In the following, we will illustrate the WHOSVR based reconstruction method for snapshot hyperspectral imaging.
\subsection{ Reconstruction Method}
Based on the analysis above, a general reconstruction formulation for snapshot hyperspectral imaging is proposed:

\begin{small}
	\begin{equation}
	%\begin{flalign}
	\begin{split}
	%{\bm{\mathcal{X}}} =
	\mathop {\min }\limits_{\bm{\mathcal{F}},\bm{\mathcal{G}}_{l}} \;&\frac{1}{2}\left\| {{\bm{Y}} - \bm{\Phi}(\bm{\mathcal{F}})} \right\|_F^2 +
	\sum\nolimits_{l = 1}^L \big (
	\\& \tau  \left\| {{{\bf{R}}_l(\bm{\mathcal{F}})} - \bm{\mathcal{G}}_l~{\times}_1~ {\bm{U}_{l,1}}~{\times}_2~{\bm{U}_{l,2}}~{\times}_3~{\bm{U}_{l,3}}} \right\|_F^2
	+  \left\| {\bf{w}}_{l} \circ {\bm{\mathcal{G}}}_{l} \right\|_1 \big ),%{{w_n}{\bf{A}}({{{\bm{\mathcal{P}}_l}_{(n)}}})})
	\end{split}
	%\end{flalign}
	\label{eqTolHSIObjFun}
	\end{equation}
\end{small}
where $L$ denotes the total number of constructed tensors. To optimize Equation \ref{eqTolHSIObjFun}, we adopt an alternating minimization scheme to split it into two finer subproblems: updating core tensor $\bm{\mathcal{G}}_l$ and updating the whole HSI $\bm{\mathcal{F}}$.
\subsubsection{Updating Core Tensor $\bm{\mathcal{G}}_l$}
By fixing HSI $\bm{\mathcal{F}}$, we can estimate each low-rank tensor $\bm{\mathcal{G}}_l$ independently by solving the following reformulated equation:
\begin{equation}
%\begin{flalign}
%{\bm{\mathcal{X}}} =
\mathop {\min }\limits_{\bm{\mathcal{G}}_{l}}  \tau   \left\| {{{\bf{R}}_l(\bm{\mathcal{F}})} - \bm{\mathcal{G}}_l~{\times}_1~ {\bm{U}_{l,1}}~{\times}_2~{\bm{U}_{l,2}}~{\times}_3~{\bm{U}_{l,3}}} \right\|_F^2
+  \left\| {\bf{w}}_l \circ \bm{\mathcal{G}}_l \right\|_1,%{{w_n}{\bf{A}}({{{\bm{\mathcal{P}}_l}_{(n)}}})})
%\end{flalign}
\label{eqSubCTObjFun}
\end{equation}

Given ${{\bf{R}}_l(\bm{\mathcal{F}})} ={\hat{\bm{\mathcal{G}}}_l}~{\times}_1~ {\bm{U}_{l,1}}~{\times}_2~{\bm{U}_{l,2}}~{\times}_3~{\bm{U}_{l,3}}$ be the Tucker decomposition of ${{\bf{R}}_l(\bm{\mathcal{F}})}$, the solution of $\bm{\mathcal{G}}_l$ in Equation~\ref{eqSubCTObjFun} is:
\begin{equation}
{g_n}   = \text{max}({\hat{g_n}}-\frac{w_n}{2\tau},0),
\label{eqSolCTObjFun}
\end{equation}
where $\hat{g_n}$ is the element of $\hat{\bm{\mathcal{G}}}_l$.
\SynErrorMap

\subsubsection{Updating Whole HSI $\bm{\mathcal{F}}$}
Once we obtain the core tensor $\bm{\mathcal{G}}_l$, the whole HSI $\bm{\mathcal{F}}$ can be updated by solving the following problem:
\begin{small}
\begin{equation}
\begin{split}
&\qquad\mathop {\min }\limits_{\bm{\mathcal{F}}} \;\frac{1}{2}\left\| {{\bm{Y}} - \bm{\Phi}(\bm{\mathcal{F}})} \right\|_F^2 +\\& 
\sum\nolimits_{l = 1}^L \tau   \left\| {{{\bf{R}}_l(\bm{\mathcal{F}})} - \bm{\mathcal{G}}_l{\times}_1 {\bm{U}_{l,1}}{\times}_2{\bm{U}_{l,2}}{\times}_3{\bm{U}_{l,3}}} \right\|_F^2.
\label{eqSubHSIObjFun}
\end{split}
\end{equation}
\end{small}
Equation~\ref{eqSubHSIObjFun} is a quadratic optimization problem, so $\bm{\mathcal{F}}$ admits a straightforward least-square solution:
\begin{small}
\begin{equation}
\begin{aligned}
\bm{\mathcal{F}} = \big{(} {{\bm{\Phi} ^T}\bm{\Phi}  + 2\tau \sum\nolimits_l {{{ {{{\bf{R}}_l^T}}}}{{\bf{R}}_l}} }\big{)} ^{{\rm{ - }}1}} \big{(} {{\bm{\Phi} ^T}(\bm{Y}) \\+ 2\tau \sum\nolimits_{l=1}^L {{{ {{{\bf{R}}_l^T}} }}(\bm{\mathcal{G}}_l~{\times}_1~ {\bm{U}_{l,1}}~{\times}_2~{\bm{U}_{l,2}}~{\times}_3~{\bm{U}_{l,3}})}  \big{)}.
\label{eqHSISolution}
\end{aligned}
\end{equation}
\end{small}

In practice, Equation~\ref{eqHSISolution} can be solved by the conjugate gradient algorithm~\cite{MR0060307}. 

\SynSpectra
\TabSynDCCHI
%\TabSynCAVECASSI
\section{Experiments}
In this section, we conduct experiments on two representative snapshot hyperspectral imaging systems, i.e., CASSI and DCCHI, to verify the performance of our method.
\subsection{Implementation Details}
We generate the mask of coded aperture in CASSI as a random Bernoulli matrix with $p=0.5$. The dispersion of the Amici prism obeys a linear distribution across the wavelength dimension. The Columbia dataset, which contains 32 various real-world objects from 400nm to 700nm (31 bands with 10nm interval), are  used as synthetic data. In our experiment, the resolution of all tested images is cropped into $256 \times 256$ at the center region cross full bands. Meanwhile, we use 22 HSIs for training and 10 HSIs for testing.

Our algorithm is compared with 5 prior knowledge based methods, i.e., TV regularization
solved by TwIST~\cite{Bioucas2007A}, 3D sparse reconstruction (3DSR)~\cite{Lin2014Spatial}, approximate message passing (AMP)~\cite{7328255}, nonlocal sparse representation (NSR)~\cite{7676344} and low-rank matrix approximation (LRMA)~\cite{fu2016exploiting} and 5 deep learning based methods, i.e., AE~\cite{choi2017high}, HSCNN~\cite{xiong2017hscnn}, ISTA-Net (shorten as ISTA)~\cite{zhang2018ista}, HRNet~\cite{wang2018hyperreconnet} and SRP\cite{wang2019hyperspectral}.

The parameters for our method are set as following. The penalty factor $\tau=1$, the constant $c=0.0055$, and the cubic spatial size $s=5$ with step length of 4. We search $k=45$ nonlocal similar patches within a [-20, 20] window. We set the max iteration number of the alternating
minimization scheme as 600 for the stop criterion. For the competitive methods we make great efforts to achieve the best results for all the competing methods according to their publication and released codes. We execute our experiments on a platform of the Windows 10 64-bit system with I7 6700 and 64GB RAM.

For quantitative evaluation, four image quality indexes are employed, including peek signal-to-noise ratio (PSNR), structure similarity (SSIM)~\cite{1284395}, erreur relative globale adimensionnelle de synth\`ese (ERGAS)\cite{wald2002data} and root mean square error (RMSE). PSNR and SSIM measure the visual quality and  the structure similarity, respectively. ERGAS and RMSE measure the spectral fidelity.  Generally, a bigger PSNR and SSIM and a smaller ERGAS and RMSE suggest a better reconstruction fidelity.
\subsection{Evaluation Results}

The average quantitative results of all methods on CASSI are shown in Table~\ref{tab:CASSI}. The best values for each index are highlighted in bold. 
We can see that our method
obtains remarkable promotion in PSNR, SSIM, ERGAS and
RMSE compared with other methods. Specifically, our method produces noticeable quality promotion compared with TV, AMP, 3DSR and NSR. It indicates that low rank prior can recover more structure information than sparse prior. The promotion upon LRMA demonstrates that the high-dimensional tensor is more powerful than matrix to exploit the intrinsic nature of HSI. Meanwhile, promotion upon the deep learning based methods indicates that the tensor based optimization can more faithfully account for the high-dimensionality structure. For a visual comparison, we convert HSI into RGB image using the CIE color mapping function. The synthetic RGB images of two representative scenes, i.e., \emph{chart and stuffed toy} and \emph{stuffed toys}, are shown in Fig.~\ref{fig:SynCASSI}. We can see that our method can produce not only clearer spatial textures but also higher spectral accuracy.

Next, we evaluate the reconstruction performance on DCCHI. It should be noted that the mismatch in the two branches of DCCHI results in that block layout based learning methods can not be implemented. So here we compare our method with the prior regularization based methods. The average quantitative results are presented in Table~\ref{tab:DCCHI}. It shows that our method can produce the best quantitative performance. We then show the average absolute error
maps between the ground truth and restored results across
spectra in Fig.~\ref{fig:ErrorMap}. It can be seen that the results produced
by our method are closer to the ground truth compared with
other methods, which verifies that our method obtains higher
spatial accuracy. 
Futher, we show the average absolute error curves between the ground truth and reconstructed results across spectra in Fig.~\ref{fig:SynSpectra}. It demonstrates that our method obtains higher spectral accuracy.

\section{Conclusion}
In this paper, we proposed a novel and general reconstruction method with low-rank tensor recovery based on WHOSVR for snapshot hyperspectral imaging. We introduce 3D tensors to exploit the high-order nature of HSI, including spatial self-similarity, spectral correlation and spatial-spectral joint correlation. Specifically, we first construct a 3D tensor for each exemplar cubic patch of HSI. We then proposed to utilize WHOSVR to characterize the high correlation in each mode of the formulated tensors. Finally, an iterative optimization algorithm based on WHOSVR was developed to finish high-accuracy HSI reconstruction. Through experiments implemented on two representative systems verified that our method outperforms state-of-the-art methods.

\section*{Acknowledgment}
This work is supported by the National Natural Science Foundation of China under Grant No.~62072038 and No.~61922014.

\bibliographystyle{unsrt}
\bibliography{sample-base}

\begin{thebibliography}{10}

\bibitem{Min20123D}
H.~Kim Min, Todd~Alan Harvey, David~S. Kittle, Holly Rushmeier, Julie Dorsey,
  Richard~O. Prum, and David~J. Brady.
\newblock 3d imaging spectroscopy for measuring hyperspectral patterns on solid
  objects.
\newblock {\em ACM Transactions on Graphics}, 31(4):1--11, 2012.

\bibitem{tarabalka2009segmentation}
Yuliya Tarabalka, Jocelyn Chanussot, and J{\'o}n~Atli Benediktsson.
\newblock Segmentation and classification of hyperspectral images using minimum
  spanning forest grown from automatically selected markers.
\newblock {\em IEEE Transactions on Systems, Man, and Cybernetics, Part B
  (Cybernetics)}, 40(5):1267--1279, 2009.

\bibitem{Pan2003Face}
Zhihong Pan, Glenn Healey, Manish Prasad, and Bruce Tromberg.
\newblock Face recognition in hyperspectral images.
\newblock {\em IEEE Transactions on Pattern Analysis and Machine Intelligence},
  25(12):1552--1560, 2003.

\bibitem{Nguyen2010Tracking}
Hien Van~Nguyen, Amit Banerjee, and Rama Chellappa.
\newblock Tracking via object reflectance using a hyperspectral video camera.
\newblock In {\em IEEE Conference on Computer Vision and Pattern Recognition
  Workshops}, pages 44--51, 2010.

\bibitem{porter1987system}
Gregg Vane, Robert~O Green, Thomas~G Chrien, Harry~T Enmark, Earl~G Hansen, and
  Wallace~M Porter.
\newblock A system overview of the airborne visible/infrared imaging
  spectrometer (aviris).
\newblock In {\em Imaging Spectroscopy II}, volume 834, pages 22--32, 1987.

\bibitem{basedow1995hydice}
Robert~W Basedow, Dwayne~C Carmer, and Mark~E Anderson.
\newblock Hydice system: Implementation and performance.
\newblock In {\em Imaging Spectrometry}, volume 2480, pages 258--268, 1995.

\bibitem{yamaguchi2006high}
Masahiro Yamaguchi, Hideaki Haneishi, Hiroyuki Fukuda, Junko Kishimoto, Hiroshi
  Kanazawa, Masaru Tsuchida, Ryo Iwama, and Nagaaki Ohyama.
\newblock High-fidelity video and still-image communication based on spectral
  information: Natural vision system and its applications.
\newblock In {\em Spectral Imaging: Eighth International Symposium on
  Multispectral Color Science}, volume 6062, page 60620G, 2006.

\bibitem{schechner2002generalized}
Yoav~Y. Schechner and Shree~K. Nayar.
\newblock Generalized mosaicing: Wide field of view multispectral imaging.
\newblock {\em IEEE Transactions on Pattern Analysis and Machine Intelligence},
  24(10):1334--1348, 2002.

\bibitem{Baraniuk2007Compressive}
Richard~G Baraniuk.
\newblock Compressive sensing [lecture notes].
\newblock {\em IEEE Signal Processing Magazine}, 24(4):118--121, 2007.

\bibitem{Cand2006Compressive}
Emmanuel~J Candes.
\newblock Compressive sampling.
\newblock In {\em Proceedings of the international congress of mathematicians},
  volume~3, pages 1433--1452, 2006.

\bibitem{Arce2014Compressive}
Gonzalo~R Arce, David~J Brady, Lawrence Carin, Henry Arguello, and David~S
  Kittle.
\newblock Compressive coded aperture spectral imaging: An introduction.
\newblock {\em IEEE Signal Processing Magazine}, 31(1):105--115, 2014.

\bibitem{Wagadarikar2008Single}
Ashwin Wagadarikar, Renu John, Rebecca Willett, and David Brady.
\newblock Single disperser design for coded aperture snapshot spectral imaging.
\newblock {\em Applied Optics}, 47(10):B44--B51, 2008.

\bibitem{Wang2015Dual}
Lizhi Wang, Zhiwei Xiong, Dahua Gao, Guangming Shi, and Feng Wu.
\newblock Dual-camera design for coded aperture snapshot spectral imaging.
\newblock {\em Applied Optics}, 54(4):848--58, 2015.

\bibitem{Wang2015High}
Lizhi Wang, Zhiwei Xiong, Dahua Gao, Guangming Shi, Wenjun Zeng, and Feng Wu.
\newblock High-speed hyperspectral video acquisition with a dual-camera
  architecture.
\newblock In {\em IEEE Conference on Computer Vision and Pattern Recognition},
  pages 4942--4950, 2015.

\bibitem{zhang2018fast}
Shipeng Zhang, Huang Hua, and Ying Fu.
\newblock Fast parallel implementation of dual-camera compressive hyperspectral
  imaging system.
\newblock {\em IEEE Transactions on Circuits and Systems for Video Technology},
  2018.

\bibitem{Figueiredo2007Gradient}
M{\'a}rio~AT Figueiredo, Robert~D Nowak, and Stephen~J Wright.
\newblock Gradient projection for sparse reconstruction: Application to
  compressed sensing and other inverse problems.
\newblock {\em IEEE Journal of selected topics in signal processing},
  1(4):586--597, 2007.

\bibitem{7328255}
Jin Tan, Yanting Ma, Hoover Rueda, Dror Baron, and Gonzalo~R Arce.
\newblock Compressive hyperspectral imaging via approximate message passing.
\newblock {\em IEEE Journal of Selected Topics in Signal Processing},
  10(2):389--401, 2016.

\bibitem{Lin2014Spatial}
Xing Lin, Yebin Liu, Jiamin Wu, and Qionghai Dai.
\newblock Spatial-spectral encoded compressive hyperspectral imaging.
\newblock {\em ACM Transactions on Graphics}, 33(6):1--11, 2014.

\bibitem{7676344}
Lizhi Wang, Zhiwei Xiong, Guangming Shi, Feng Wu, and Wenjun Zeng.
\newblock Adaptive nonlocal sparse representation for dual-camera compressive
  hyperspectral imaging.
\newblock {\em IEEE Transactions on Pattern Analysis and Machine Intelligence},
  39(10):2104--2111, 2017.

\bibitem{fu2016exploiting}
Ying Fu, Yinqiang Zheng, Imari Sato, and Yoichi Sato.
\newblock Exploiting spectral-spatial correlation for coded hyperspectral image
  restoration.
\newblock In {\em IEEE Conference on Computer Vision and Pattern Recognition},
  pages 3727--3736, 2016.

\bibitem{liu2018rank}
Yang Liu, Xin Yuan, Jinli Suo, David Brady, and Qionghai Dai.
\newblock Rank minimization for snapshot compressive imaging.
\newblock {\em IEEE Transactions on Pattern Analysis and Machine Intelligence},
  2018.

\bibitem{Cao2011A}
Xun Cao, Hao Du, Xin Tong, Qionghai Dai, and Stephen Lin.
\newblock A prism-mask system for multispectral video acquisition.
\newblock {\em IEEE Transactions on Pattern Analysis and Machine Intelligence},
  33(12):2423--2435, 2011.

\bibitem{cao2016computational}
Xun Cao, Tao Yue, Xing Lin, Stephen Lin, Xin Yuan, Qionghai Dai, Lawrence
  Carin, and David~J Brady.
\newblock Computational snapshot multispectral cameras: toward dynamic capture
  of the spectral world.
\newblock {\em IEEE Signal Processing Magazine}, 33(5):95--108, 2016.

\bibitem{gao2016review}
Liang Gao and Lihong~V Wang.
\newblock A review of snapshot multidimensional optical imaging: Measuring
  photon tags in parallel.
\newblock {\em Physics reports}, 616:1--37, 2016.

\bibitem{Kittle2010Multiframe}
David Kittle, Kerkil Choi, Ashwin Wagadarikar, and David~J Brady.
\newblock Multiframe image estimation for coded aperture snapshot spectral
  imagers.
\newblock {\em Applied Optics}, 49(36):6824, 2010.

\bibitem{wu2011development}
Yuehao Wu, Iftekhar~O Mirza, Gonzalo~R Arce, and Dennis~W Prather.
\newblock Development of a digital-micromirror-device-based multishot snapshot
  spectral imaging system.
\newblock {\em Optics Letters}, 36(14):2692--2694, 2011.

\bibitem{Wagadarikar2008Spectral}
Ashwin~A Wagadarikar, Nikos~P Pitsianis, Xiaobai Sun, and David~J Brady.
\newblock Spectral image estimation for coded aperture snapshot spectral
  imagers.
\newblock In {\em Image Reconstruction from Incomplete Data V}, volume 7076,
  page 707602, 2008.

\bibitem{8529273}
Shipeng Zhang, Huang Hua, and Ying Fu.
\newblock Fast parallel implementation of dual-camera compressive hyperspectral
  imaging system.
\newblock {\em IEEE Transactions on Circuits and Systems for Video Technology},
  pages 1--1, 2018.

\bibitem{zhang2018gpu}
Shipeng Zhang, Lizhi Wang, Ying Fu, and Hua Huang.
\newblock Gpu assisted towards real-time reconstruction for dual-camera
  compressive hyperspectral imaging.
\newblock In {\em Pacific Rim Conference on Multimedia}, pages 711--720.
  Springer, 2018.

\bibitem{sun2016deep}
Jian Sun, Huibin Li, and Zongben Xu.
\newblock Deep admm-net for compressive sensing mri.
\newblock In {\em Advances in Neural Information Processing Systems}, pages
  10--18, 2016.

\bibitem{zhang2018ista}
Jian Zhang and Bernard Ghanem.
\newblock Ista-net: Interpretable optimization-inspired deep network for image
  compressive sensing.
\newblock In {\em IEEE Conference on Computer Vision and Pattern Recognition},
  pages 1828--1837, 2018.

\bibitem{choi2017high}
Inchang Choi, Daniel~S Jeon, Giljoo Nam, Diego Gutierrez, and Min~H Kim.
\newblock High-quality hyperspectral reconstruction using a spectral prior.
\newblock {\em ACM Transactions on Graphics}, 36(6):218, 2017.

\bibitem{xiong2017hscnn}
Zhiwei Xiong, Zhan Shi, Huiqun Li, Lizhi Wang, Dong Liu, and Feng Wu.
\newblock Hscnn: Cnn-based hyperspectral image recovery from spectrally
  undersampled projections.
\newblock In {\em IEEE International Conference on Computer Vision Workshops},
  pages 518--525, 2017.

\bibitem{wang2018hyperreconnet}
Lizhi Wang, Tao Zhang, Ying Fu, and Hua Huang.
\newblock Hyperreconnet: Joint coded aperture optimization and image
  reconstruction for compressive hyperspectral imaging.
\newblock {\em IEEE Transactions on Imaging Processing}, 28(5):2257--2270,
  2018.

\bibitem{wang2019hyperspectral}
Lizhi Wang, Chen Sun, Ying Fu, Min~H Kim, and Hua Huang.
\newblock Hyperspectral image reconstruction using a deep spatial-spectral
  prior.
\newblock In {\em IEEE Conference on Computer Vision and Pattern Recognition},
  pages 8032--8041, 2019.

\bibitem{tucker1966some}
Ledyard~R Tucker.
\newblock Some mathematical notes on three-mode factor analysis.
\newblock {\em Psychometrika}, 31(3):279--311, 1966.

\bibitem{MR0060307}
Magnus~R. Hestenes and Eduard Stiefel.
\newblock Methods of conjugate gradients for solving linear systems.
\newblock {\em J. Research Nat. Bur. Standards}, 49:409--436 (1953), 1952.

\bibitem{Bioucas2007A}
Jos{\'e}~M Bioucas-Dias and M{\'a}rio~AT Figueiredo.
\newblock A new twist: Two-step iterative shrinkage/thresholding algorithms for
  image restoration.
\newblock {\em IEEE Transactions on Image Processing}, 16(12):2992--3004, 2007.

\bibitem{1284395}
Zhou Wang, A.~C. Bovik, H.~R. Sheikh, and E.~P. Simoncelli.
\newblock Image quality assessment: from error visibility to structural
  similarity.
\newblock {\em IEEE Transactions on Image Processing}, 13(4):600--612, 2004.

\bibitem{wald2002data}
Lucien Wald.
\newblock {\em Data fusion: definitions and architectures: fusion of images of
  different spatial resolutions}.
\newblock Presses des MINES, 2002.

\end{thebibliography}

\end{document}